\documentclass[12pt,aps,prb,preprint,showpacs,showkeys]{revtex4}

\usepackage{graphicx}
\usepackage{subfigure}
\usepackage{amsmath}
\usepackage{amssymb}
\usepackage{bm}

\begin{document}

\title{Frustrated ground states of a generalized XY model and their mapping to non-magnetic structural analogues}
\author{Milan \v{Z}ukovi\v{c}}
 \email{milan.zukovic@upjs.sk}
 \affiliation{Institute of Physics, Faculty of Science, P. J. \v{S}af\'arik University, Park Angelinum 9, 041 54 Ko\v{s}ice, Slovakia}
\date{\today}

\begin{abstract}
Ground state phases of a generalized XY model with magnetic and generalized nematic couplings on a non-bipartite triangular lattice are investigated in the exchange interactions parameter space. We demonstrate that the model displays a number of ordered and quasi-ordered phases as a result of geometrical frustration and/or competition between the magnetic and generalized nematic interactions. The nature and the extent of the respective phases depend on the parameter $q$, characterizing the higher-order harmonics term in the Hamiltonian. Motivated by the recent discovery of the experimental realization of the model with $q=2$ in the seemingly unrelated field of the systems chemistry [A.B. Cairns et al., Nature Chemistry 8, 442 (2016)], the model for $q>2$ is discussed in the context of the prediction of structural phases of a class of bimetalic cyanides based on a mapping between the two systems.
\end{abstract}

\pacs{05.10.Ln, 05.50.+q, 64.60.De, 75.10.Hk, 75.30.Kz}

\keywords{Generalized XY model, Geometrical frustration, Triangular lattice, Ground state, Structural analogues}



\maketitle

\section{Introduction}
A standard two-dimensional continuous XY spin model is known to exhibit a Kosterlitz-Thouless phase transition, due to the vortex-antivortex (topological defects) pairs unbinding~\cite{kost73}, to a quasi-long-range-order (LRO) phase characterized by a power-law decaying correlation function.

The model with antiferromagnetic interactions on a non-bipartite, such as triangular, lattice becomes geometrically frustrated and it has been intensively studied in relation with the possibility of separate phase transitions to the vector chiral LRO and the magnetic quasi-LRO phases (spin-chirality decoupling) and the corresponding universality classes~\cite{miya84,lee86,lee98,kors02,hase05,obu12}. 

The model can be generalized by including the higher order harmonics, leading to the Hamiltonian
\begin{equation}
\label{Hamiltonian}
H=-J_1\sum_{\langle i,j \rangle}\cos(\phi_{i}-\phi_j)-J_q\sum_{\langle i,j \rangle}\cos[q(\phi_{i}-\phi_j)],
\end{equation}
where $\phi_i \in [0,2\pi]$ is an $i$-th site spin angle, $J_1$, $J_q$ are exchange interaction parameters and $\langle i,j \rangle$ denotes the sum over nearest-neighbor spins. The first term $J_1$ is a usual magnetic coupling, while the second term $J_q$ represents a generalized nematic interaction.

The model~(\ref{Hamiltonian}) with $q=2$ has been studied for the non-frustrated both $J_1$ and $J_2$ positive, i.e., ferromagnetic (FM) and nematic (N) interactions~\cite{lee85,kors85,sluc88,carp89,qi13} and more recently also for the frustrated both $J_1$ and $J_2$ negative, i.e., antiferromagnetic (AFM) and antinematic (AN) interactions~\cite{park08}. In both cases the ground states have been shown to be not affected by the presence of the nematic terms as long as the magnetic interactions are non-zero, i.e., for any finite ratio $J_2/J_1$ the ground state is FM in the former and AFM in the latter case. The model with the mixed signs of $J_1$ and $J_2$ on a bipartite (non-frustrated) square lattice has been shown to be applicable in modeling of high-temperature cuprate superconductors~\cite{hlub08,kome10}. However, as far as we are aware, the non-bipartite triangular lattice model with the frustration and/or competition inducing magnetic and nematic interactions of mixed signs has not been studied yet. Notwithstanding, the results obtained for a three-dimensional layered-triangular lattice XY model with different types of intra- and inter-layer magnetic and nematic interactions, reported in a series of papers~\cite{zuko01-02a,zuko01-02b,zuko02-03,zuko03b}, suggest the presence of some non-trivial complex ground states resulting from the intra-layer geometrical frustration and competition between the magnetic and nematic couplings.

A recent study of the model with $q>2$ and positive both the magnetic and the generalized nematic interactions by Poderoso et al.~\cite{pode11} has revealed that the increasing value of $q$ can drastically change the phase diagram topology, featuring different phases belonging to a variety of universality classes. This finding is rather surprising, as it points to a significant lack of universality in the systems showing the same $\phi \to \phi+2\pi$ symmetry and thus raises a more general question about the credibility of the conclusions regarding the thermodynamic behavior of the system drawn from a coarse-grained Hamiltonian. Besides this intriguing theoretical aspect, the above generalized model with $q=2$ and frustrated interactions has been demonstrated to be applicable to modeling such diverse phenomena as DNA packing~\cite{grason08} or very recently structural phases of certain cyanide polymers~\cite{cairns16,clark16}. It is interesting that this frustrated spin model, studied theoretically over decades basically as an unrealizable toy model in the field of magnetism, found its first experimental realizations as a structural analogue of the systems in seemingly completely unrelated fields.  

Motivated by the above raised theoretical questions as well as further possible experimental realizations, in the present study we investigate ground-state and near-ground-state properties of the generalized XY model in a wide space of the parameters and discuss its application to the prediction of structural phases of a class of bimetalic cyanides based on an appropriate mapping between the two systems.   

\section{Model and Methods}

In the following we consider the model~(\ref{Hamiltonian}) for general $q$ and the interaction parameters $J_1, J_q \in [-1,1]$ in the form $J_1=\cos(\theta)$, $J_q=\sin(\theta)$, with $\theta \in [0,2\pi)$, in order to cover all the possible signs and strength ratios of the interactions.
\subsection{Global optimization}
\begin{figure}[t!]
\centering
\includegraphics[scale=0.5,clip]{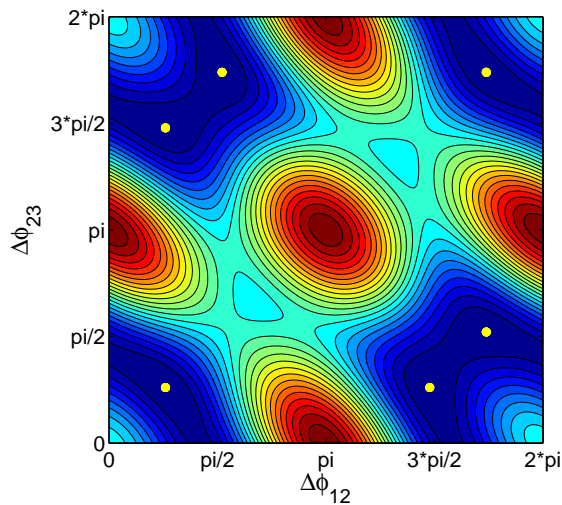}
\caption{(Color online) The objective function~(\ref{OF}), for $q=2$ and $\theta=7\pi/4$. The yellow circles denote global minima.}\label{fig:ene_q2_IV}
\end{figure}

The ground states of the model can be obtained by finding global minima of the energy functional~(\ref{Hamiltonian}) in the phase space. Assuming spin uniformity on each of the three sublattices of the triangular lattice, one basically needs to minimize the objective function
\begin{equation}
\label{OF}
F(\Delta\phi_{12},\Delta\phi_{23})=-\sum_{k=1,q}J_k(\cos(k\Delta\phi_{12})+\cos(k\Delta\phi_{23})+\cos(k[\Delta\phi_{12}+\Delta\phi_{23}])),
\end{equation}
where $\Delta\phi_{ij}$ is the phase angle between the sublattices $i$ and $j$. One should keep in mind that the surface of the objective function is generally complex and multimodal, particularly in the case of the frustrated and/or the competing magnetic and generalized nematic interactions. Therefore, care should be taken in order to find a true global minimum which, moreover, may not be unique. We note that this is the reason why we opted for the global optimization of $F$, instead of solving the set of equations $\partial F/\partial \Delta\phi_{12}=0$; $\partial F/\partial \Delta\phi_{23}=0$. In Fig.~\ref{fig:ene_q2_IV} we show an example of such a case for $q=2$ and $\theta=7\pi/4$, i.e., $J_1>0,J_q<0$, with six global (stable) solutions, marked by the yellow circles. However, considering the symmetry under sublattice exchange $\Delta\phi_{12} \leftrightarrow \Delta\phi_{23}$, in fact, there are only four different solutions.

\subsection{Monte Carlo simulation}
In our Monte Carlo (MC) simulations we consider spin systems of a linear size $L$, with the periodic boundary conditions to eliminate boundary effects. We use the standard Metropolis algorithm and for thermal averaging we take $N_{MC}$ MC sweeps after discarding another $N_0=0.2\times N_{MC}$ MC sweeps for thermalization. The simulations are performed at sufficiently low temperature $T=0.05$, to approximate ground-state conditions, and for $\theta \in [0,2\pi)$, with the step $\Delta\theta=\pi/180$ , to cover the entire parameter plane $J_1-J_q$ with sufficient resolution. Considering such a low simulation temperature, in order to secure equilibrium conditions we chose relatively small but for the purpose sufficient lattice sizes of $L=12-48$ and used $N_{MC}=10^4$ MC sweeps. We calculated the following quantities: the internal energy per spin $e=\langle H \rangle/L^2$, the magnetic ($m_1$) and the generalized nematic ($m_q$) order parameters, defined by
\begin{equation}
m_k=\langle M_k \rangle/L^2=\left\langle\sqrt{3\sum_{\alpha=1}^{3}
{\bm M}_{k\alpha}^{2}}\ \right\rangle/L^2, k=1,q; \alpha=1,2,3,
\label{mk}
\end{equation}
where ${\bm M}_{k\alpha}$ is the $\alpha$-th sublattice order parameter vector, given by
\begin{equation}
{\bm M}_{k\alpha}=\left(\sum_{i \in \alpha}\cos(k\phi_{\alpha i}),\ \sum_{i \in \alpha}\sin(k\phi_{\alpha i})\right)
\label{m_sub}
\end{equation}
and the generalized (staggered) chiralities
\begin{equation}
\kappa_k=\langle K_k \rangle/L^2=\left\langle\left|\sum_{p^{+} \in \Delta}\kappa_{kp^{+}}-\sum_{p^{-} \in \nabla}\kappa_{kp^{-}}\right|\ \right\rangle/(2L^2), k=1,q,
\label{kk}
\end{equation}
where $\kappa_{kp^{+}}$ and $\kappa_{kp^{-}}$ are the local generalized chiralities for each elementary plaquette of upward and downward triangles, respectively, defined by
\begin{equation}
\kappa_{kp}=2(\sin[k(\phi_2-\phi_1)]+\sin[k(\phi_3-\phi_2)]+\sin[k(\phi_1-\phi_3)])/3\sqrt{3},
\label{k_el}
\end{equation}
where the summation runs over the three directed bonds surrounding each plaquette, $p$, and $\phi_i$, $i=1,2,3$, represent the spin angles. $\kappa_p$ is an Ising-like quantity representing the sign of rotation of the spins along the three sides of each plaquette (see Fig.~\ref{fig:sym_quad}, for $q=1$ and $\theta=3\pi/2$). Finally, the spin correlation function is obtained as
\begin{equation}
C(r)=\left\langle\sum_{i}\cos(\phi_{i}-\phi_{i+r})\right\rangle/L^2,
\label{cf}
\end{equation}
where $\phi_{i},\phi_{i+r}$ are the turn angles of the spins separated by the distance $r$. For the standard XY model, $C(r)$ is known to decay as a power law with the temperature-dependent exponent $\eta(T)$
\begin{equation}
C(r) \sim r^{-\eta(T)}.
\label{cf_pl}
\end{equation}

MC simulations serve to complement the Hamiltonian optimization results by providing additional information, such as the order parameter values, spin snapshots, as well as the spin correlation decay, which are particularly helpful in understanding the behavior in various non-trivial phases. On the other hand, the global optimization results check the consistency of the two approaches and also verify whether the MC results have been obtained under equilibrium conditions even in this generally difficult to equilibrate near-ground-state region of frustrated systems.
 
\section{Results}
\subsection{Ground states}
\subsubsection{Purely magnetic or generalized nematic interactions}

\begin{figure}[t!]
\centering
\includegraphics[scale=0.5,clip]{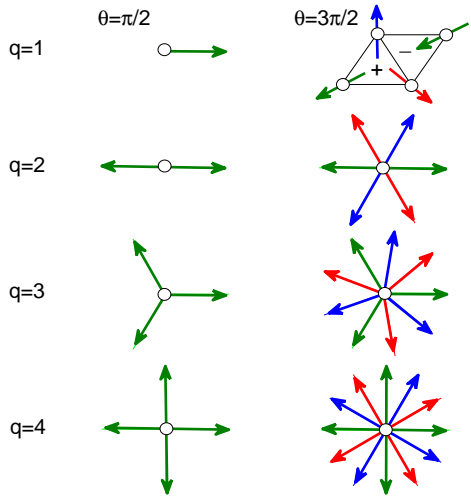}
\caption{(Color online) Schematic representation of the possible spin orientations for the models with the purely positive, i.e., $\theta=\pi/2$ (left column), and the purely negative, i.e., $\theta=3\pi/2$ (right column), magnetic ($q=1$) and generalized nematic ($q=2,3,4$) interactions. Different colors, for $\theta=3\pi/2$, represent the three sublattices of the triangular lattice and the $+$, $-$ signs, for $q=1$, represent the local chirality values.}\label{fig:sym_quad}
\end{figure} 

Ordering in the case of purely magnetic interactions, i.e., the case of $J_q=0$ with $\theta=0$ $(J_1>0)$ or $\theta=\pi$ $(J_1<0)$, is very well known. The ground state is FM in the former and coexisting AFM and Ising-like staggered chiral in the latter case. This is schematically shown in Fig.~\ref{fig:sym_quad}, for $q=1$, as a special case of the absent generalized nematic interactions. Namely, for $J_1>0$, there is a directional order among spins on the entire lattice and, for $J_1<0$, the directional order among spins on the individual sublattices (the arrows of different colors) coexists with the staggered chiral order among triangular plaquettes with opposite handedness ($+$ and $-$ signs in the upward and downward triangles, respectively).

\begin{figure}[t!]
\centering
\subfigure{\includegraphics[scale=0.5,clip]{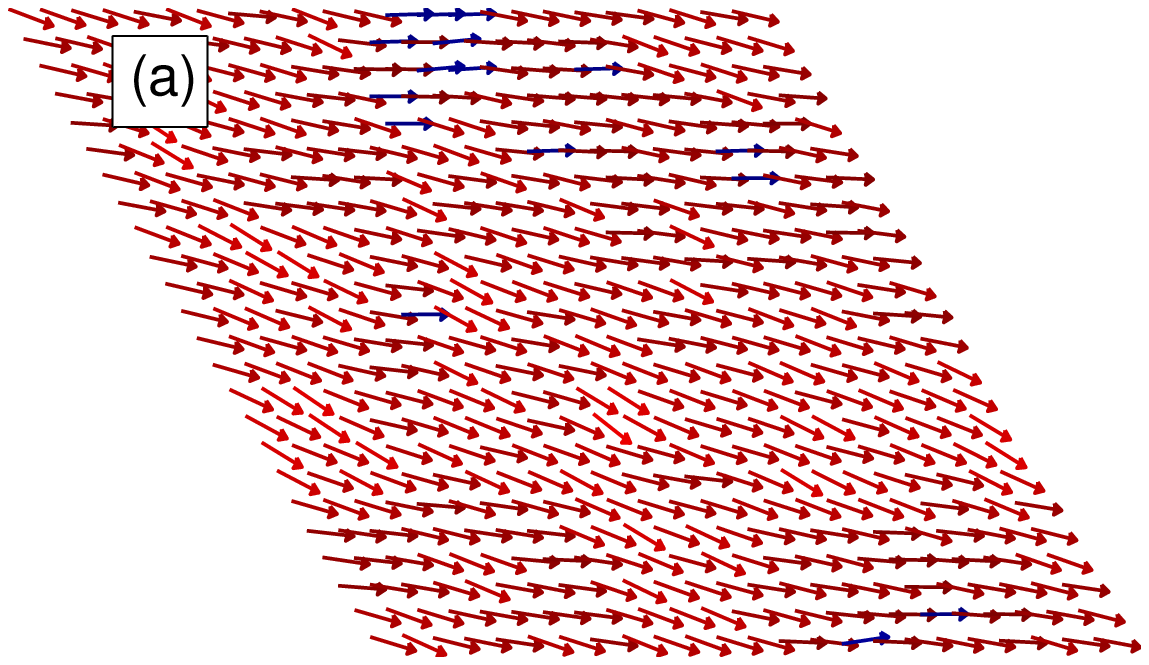}\label{fig:snap_q1_I_c}}
\subfigure{\includegraphics[scale=0.5,clip]{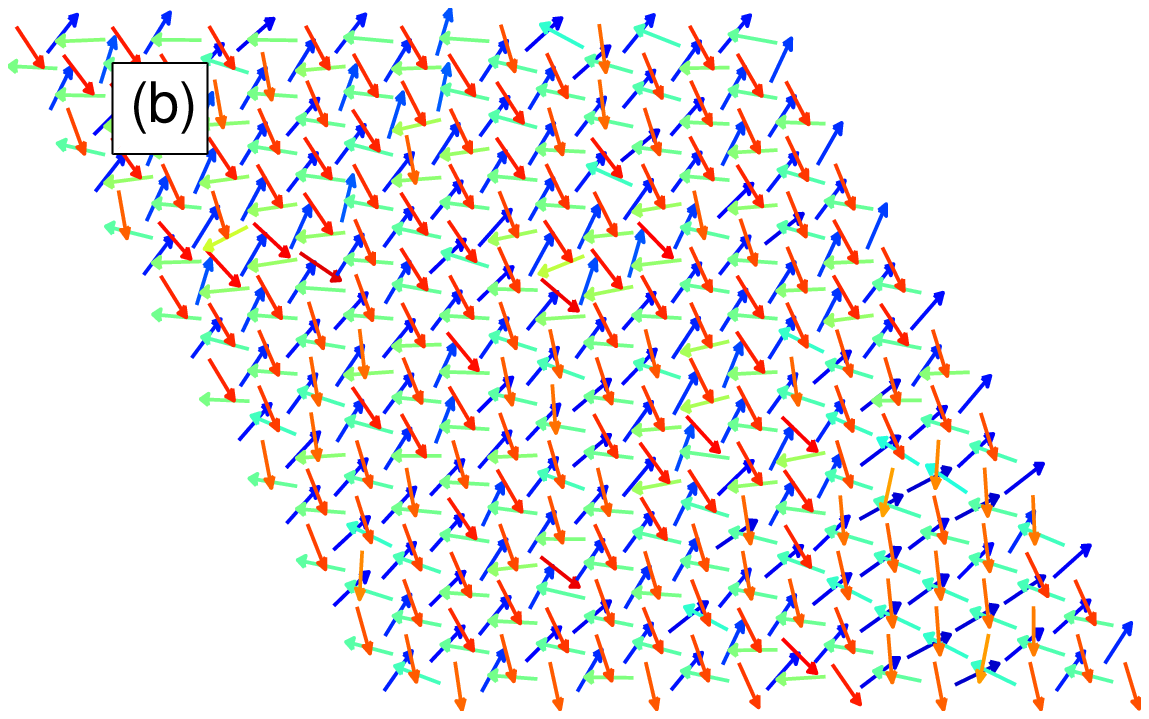}\label{fig:snap_q1_II-III_c}}\\
\subfigure{\includegraphics[scale=0.5,clip]{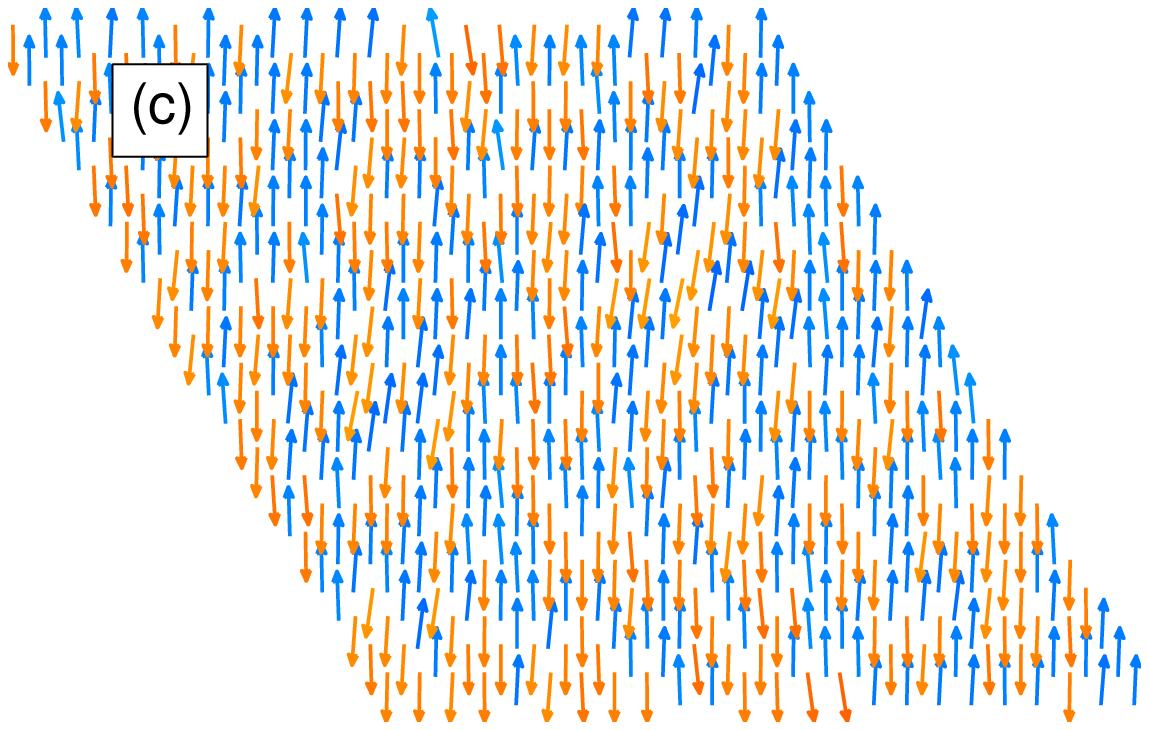}\label{fig:snap_q2_I-II_c}}
\subfigure{\includegraphics[scale=0.5,clip]{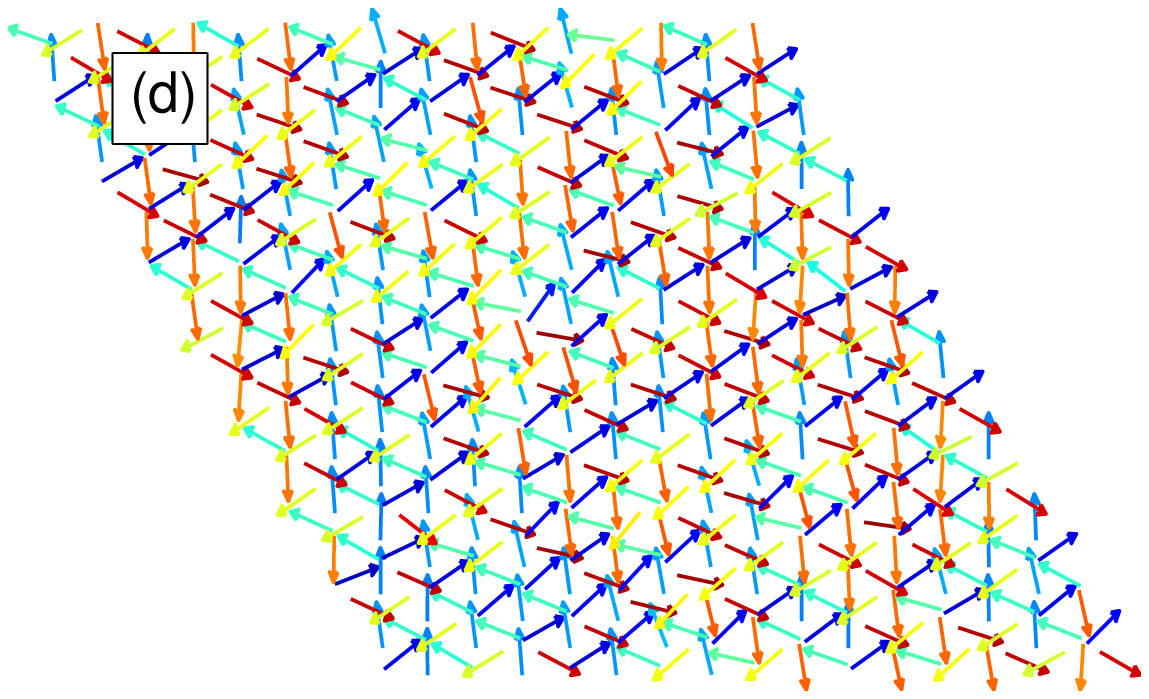}\label{fig:snap_q2_III-IV_c}}\\
\caption{(Color online) Near-ground-state ($T=0.05$) spin snapshots with the purely (a) FM and (b) AFM interactions and purely (c) N and (d) AN interactions for $q=2$.}\label{fig:snaps_pure}
\end{figure} 

On the other hand, for the model with the purely positive generalized nematic ($\theta=\pi/2$) interactions, it is easy to verify that the number of possible spin orientations in the ground state is equal to $q$, as demonstrated in the left column of Fig.~\ref{fig:sym_quad}. Similar claim can be made also for the case with the purely generalized antinematic ($\theta=3\pi/2$) interactions if one restricts the considerations to the separate sublattices. However, due to the geometrical frustration originating in the triangular lattice topology the minimum energy condition is satisfied when the spin angles on the respective sublattices are uniformly shifted with respect to each other, as illustrated in the right column of Fig.~\ref{fig:sym_quad}. Thus, the resulting number of possible spin angles in the case of the purely generalized antinematic interactions is equal to $3q$, not taking into account the rotational symmetry. Consequently, the relative turn angles between the neighboring spins from different sublattices can be defined as $\Delta\phi\in\{2i\pi/3q,i=1,\ldots,3q\}\setminus\{2j\pi/q,j=1,\ldots,q\}$. Notwithstanding, unlike in the case of the purely magnetic interactions, the minimum energy condition does not require any directional ordering among spins on the lattice for the positive $J_q$ or on the individual sublattices for the negative $J_q$. The situation is demonstrated in Fig.~\ref{fig:snaps_pure} on the MC simulation snapshots close to the ground state for the systems with the purely FM and AFM interactions, on the one hand, and the purely N and AN with $q=2$ interactions, on the other hand. 

\subsubsection{Mixed magnetic and generalized nematic interactions}
\begin{figure}[t!]
\centering
\subfigure{\includegraphics[scale=0.29,clip]{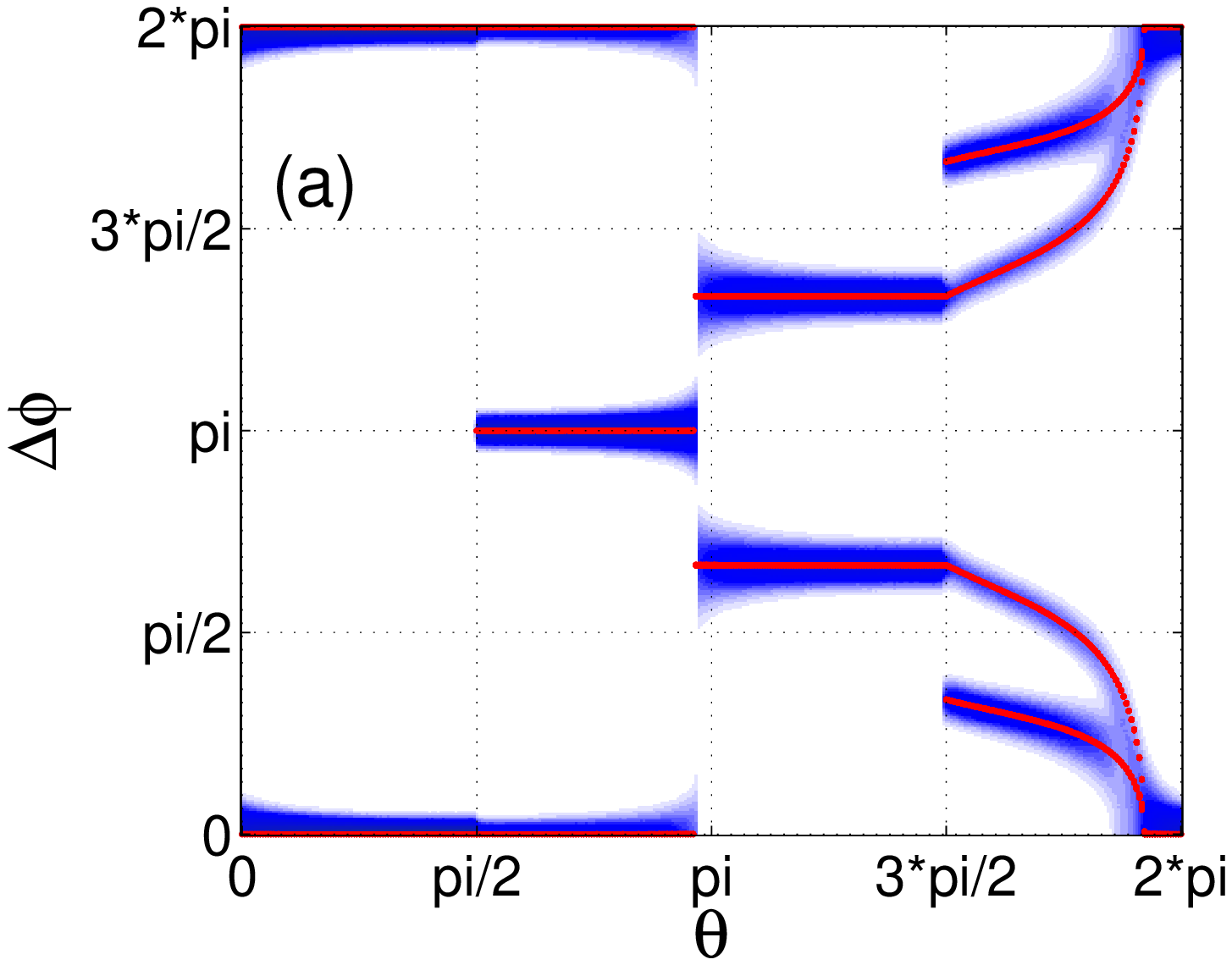}\label{fig:gs_hs_J1_J2_q2_phi}}
\subfigure{\includegraphics[scale=0.29,clip]{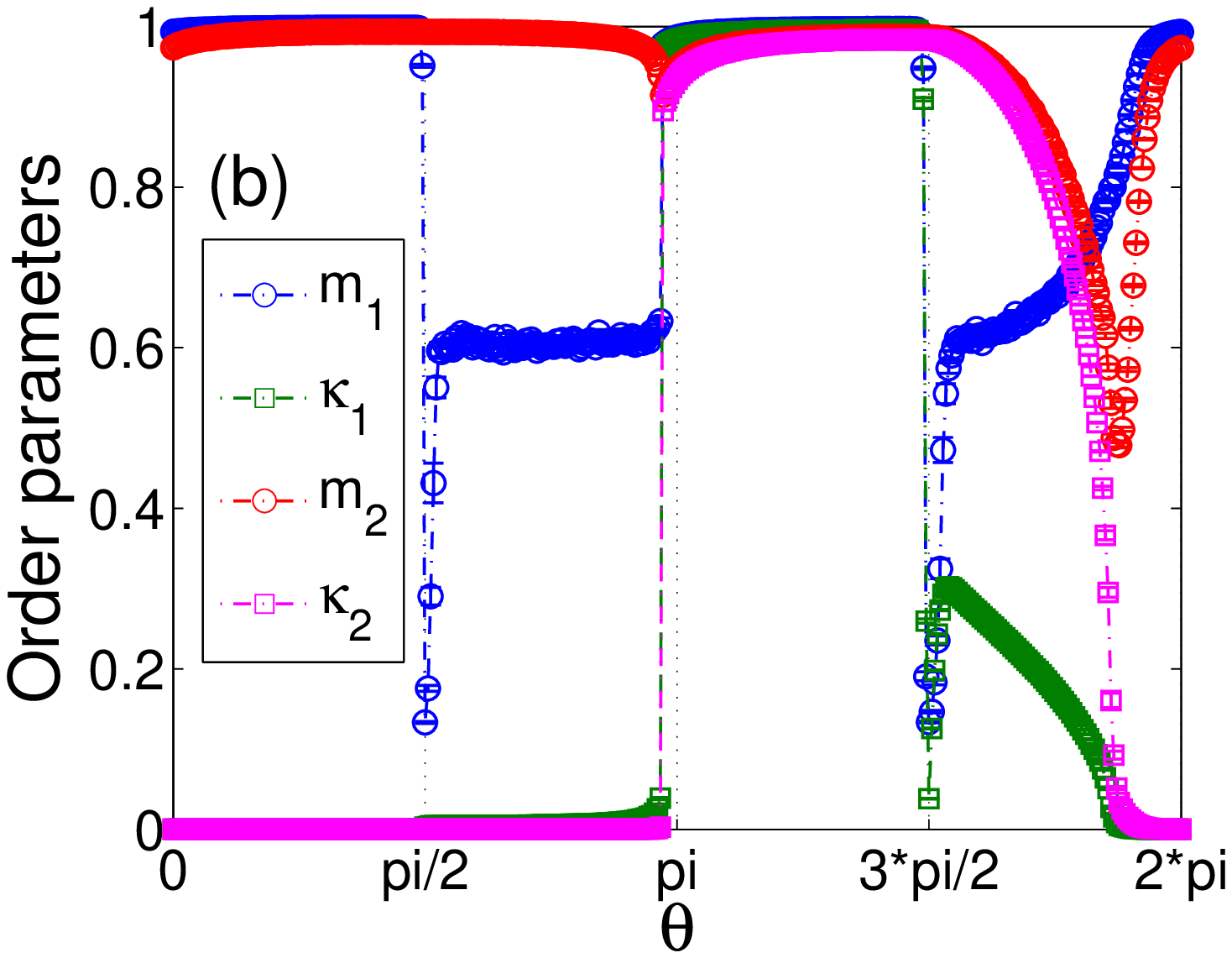}\label{fig:gs_J1_J2_q2_op}}
\subfigure{\includegraphics[scale=0.29,clip]{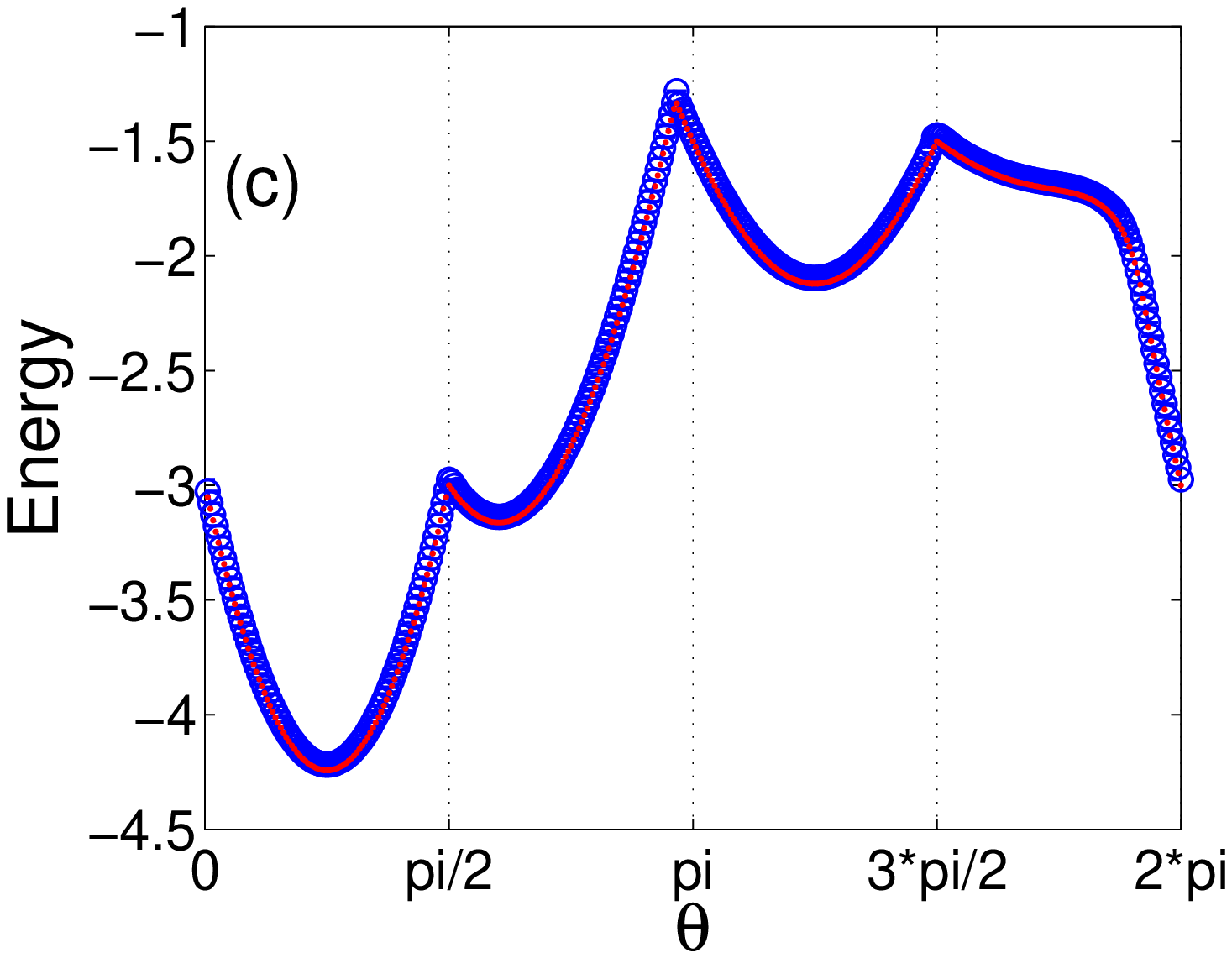}\label{fig:gs_J1_J2_q2_e}}\\ \vspace{-4mm}
\subfigure{\includegraphics[scale=0.29,clip]{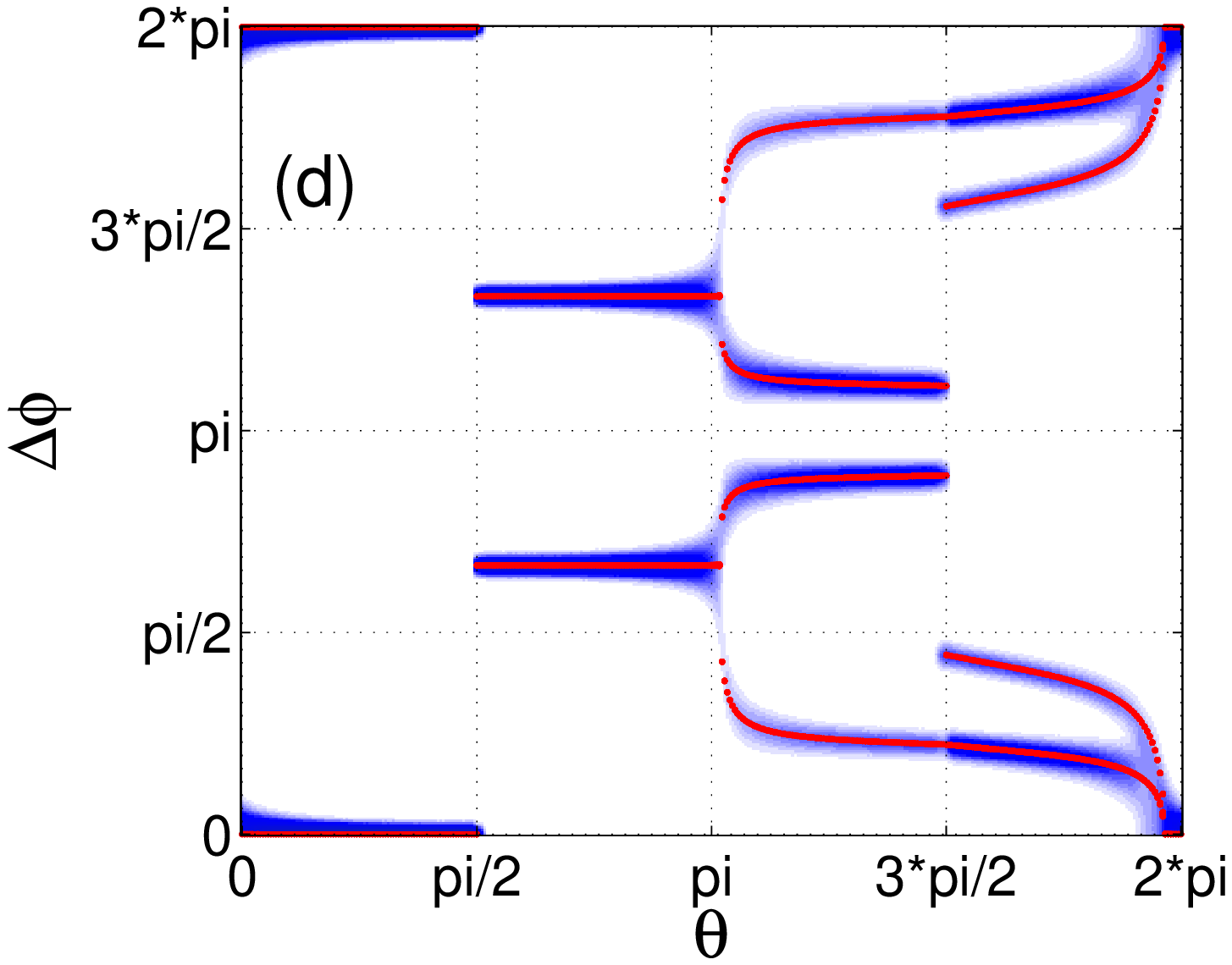}\label{fig:gs_hs_J1_J2_q3_phi}}
\subfigure{\includegraphics[scale=0.29,clip]{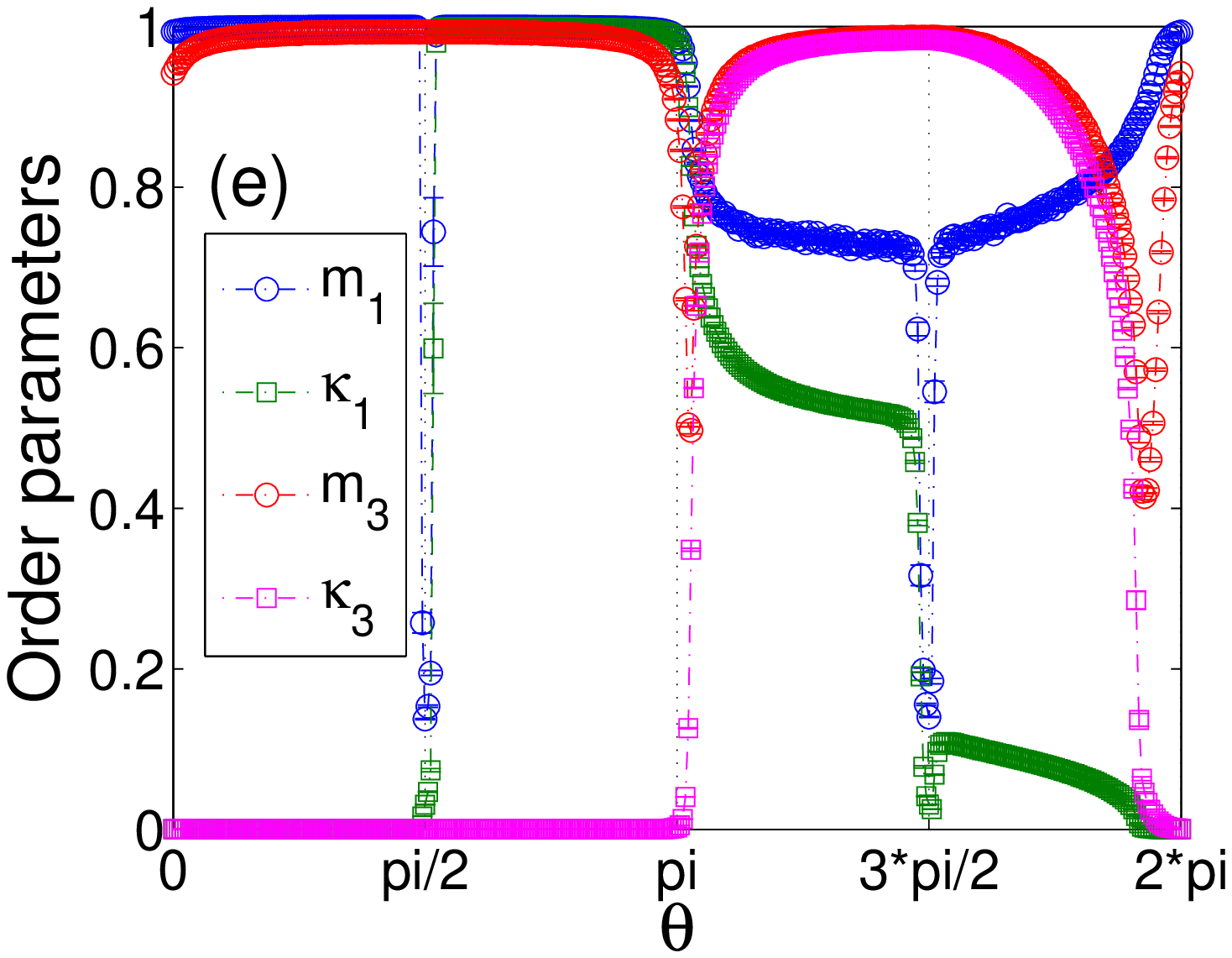}\label{fig:gs_J1_J2_q3_op}}
\subfigure{\includegraphics[scale=0.29,clip]{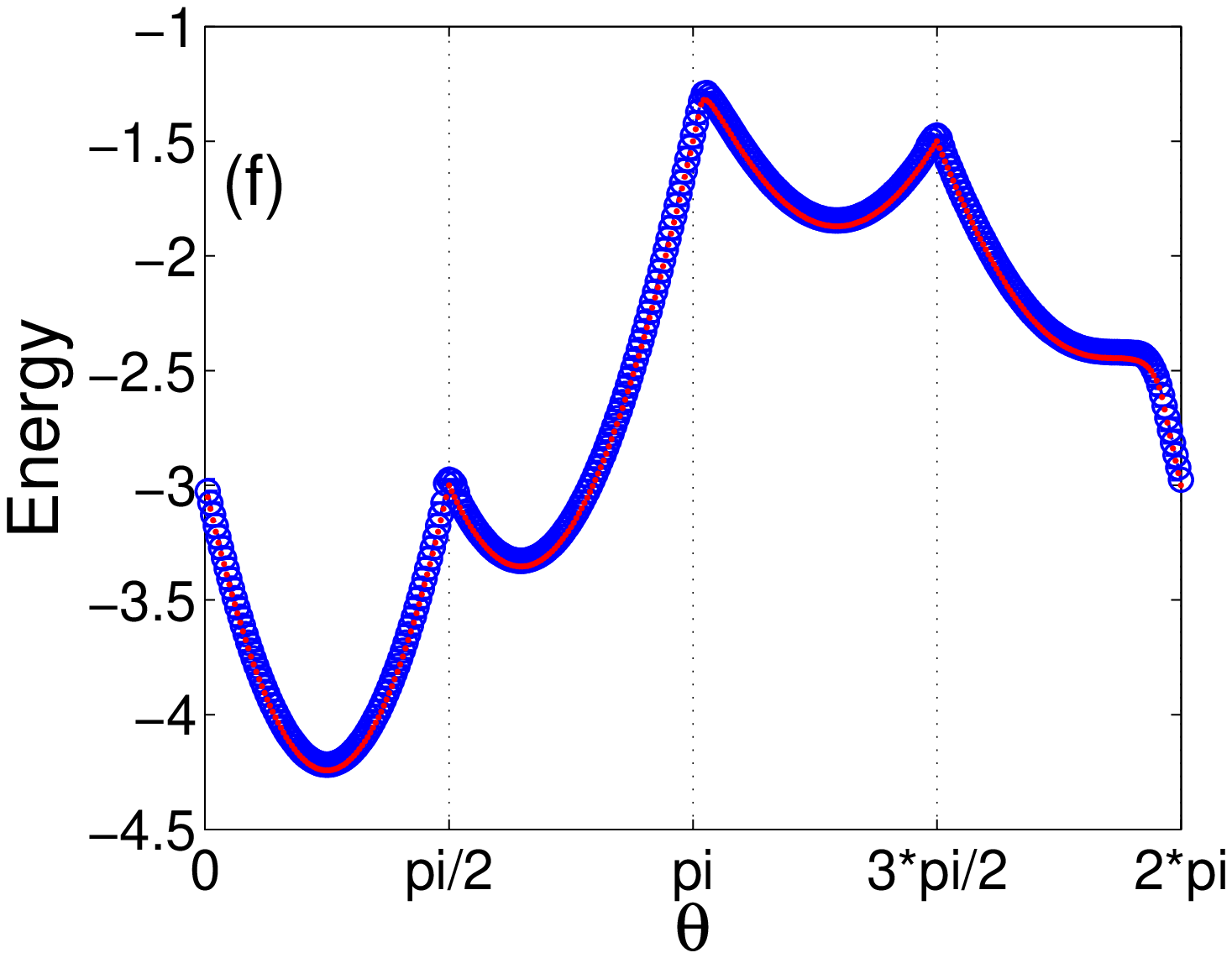}\label{fig:gs_J1_J2_q3_e}}\\ \vspace{-4mm}
\subfigure{\includegraphics[scale=0.29,clip]{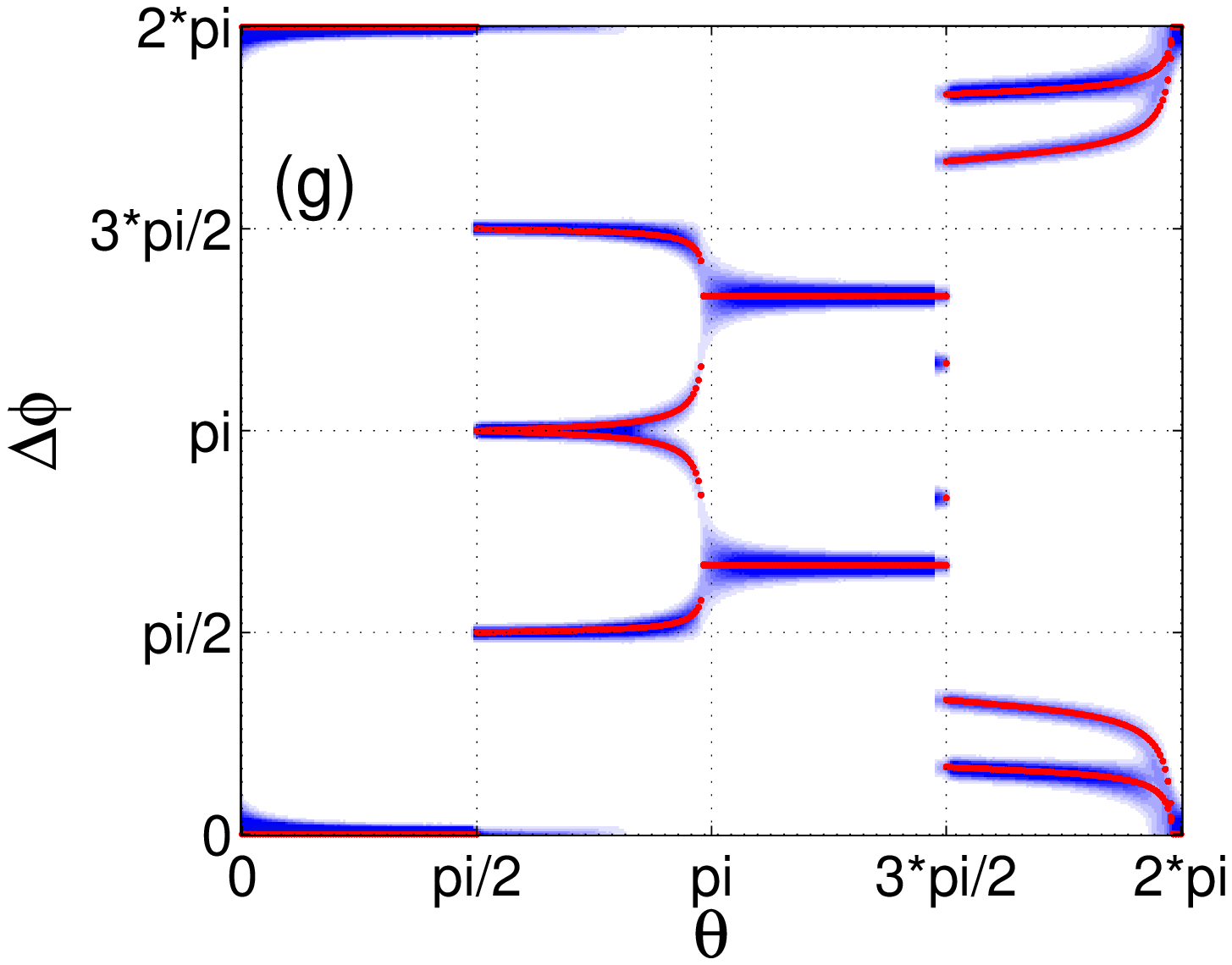}\label{fig:gs_hs_J1_J2_q4_phi}}
\subfigure{\includegraphics[scale=0.29,clip]{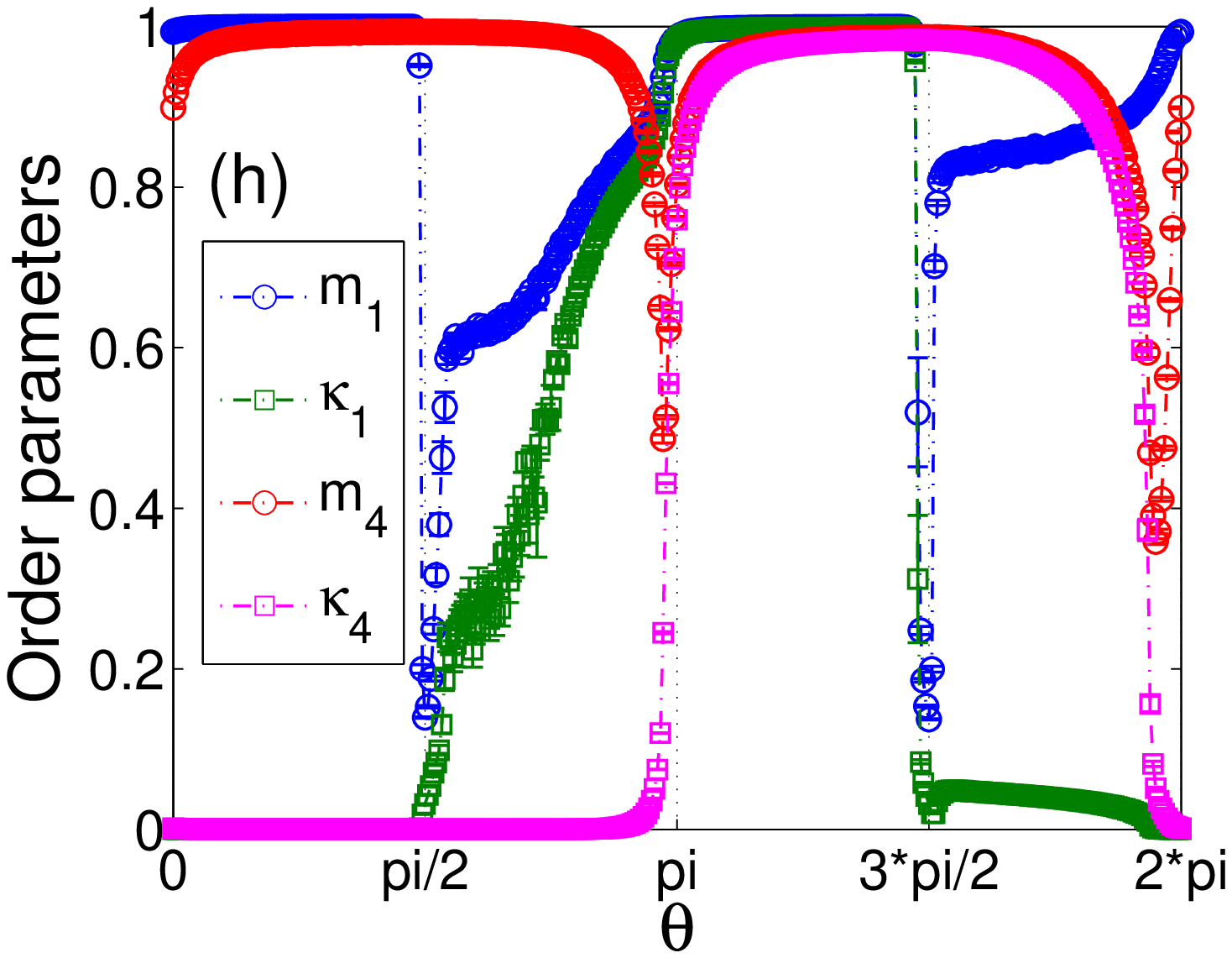}\label{fig:gs_J1_J2_q4_op}}
\subfigure{\includegraphics[scale=0.29,clip]{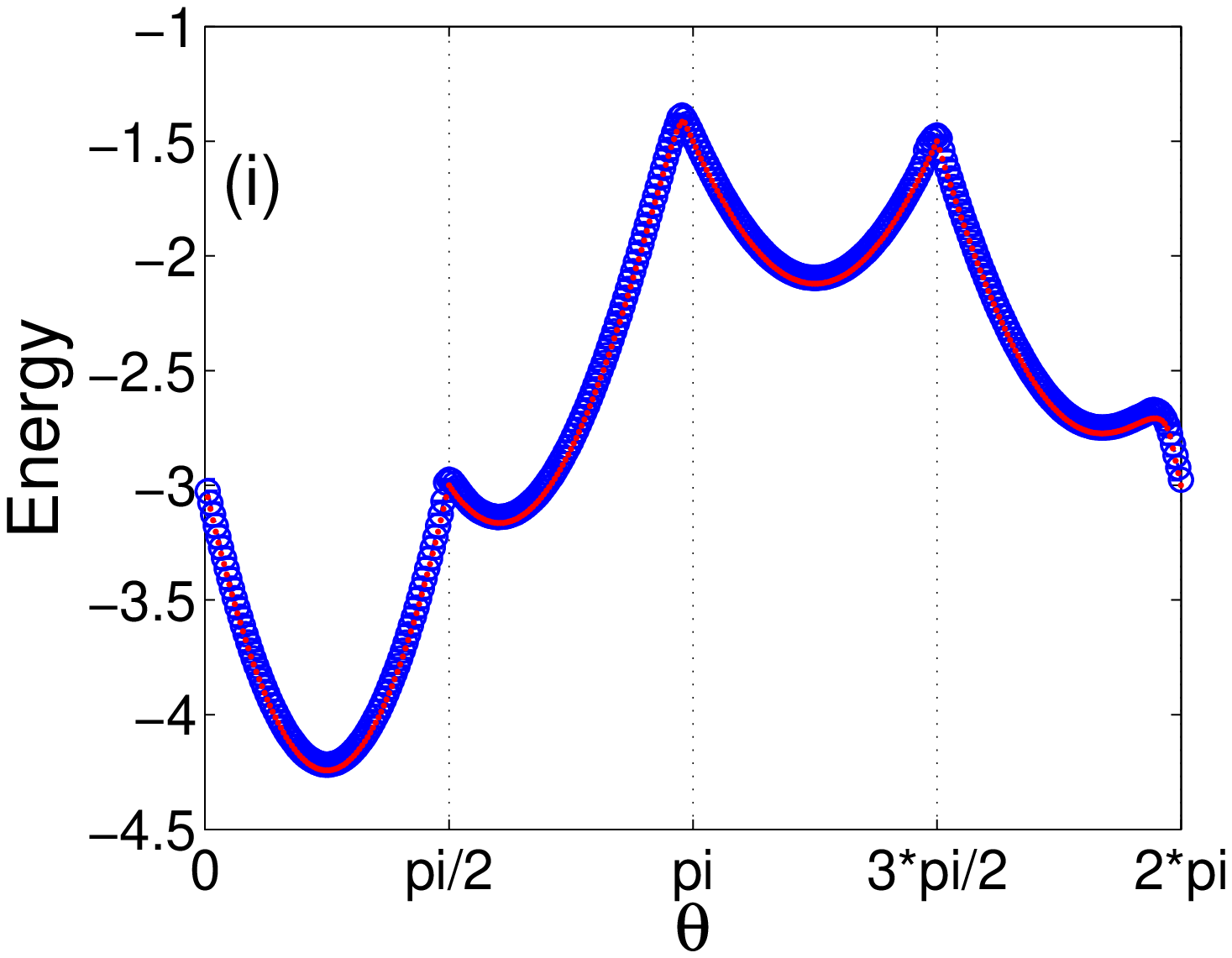}\label{fig:gs_J1_J2_q4_e}}\\ \vspace{-4mm}
\subfigure{\includegraphics[scale=0.29,clip]{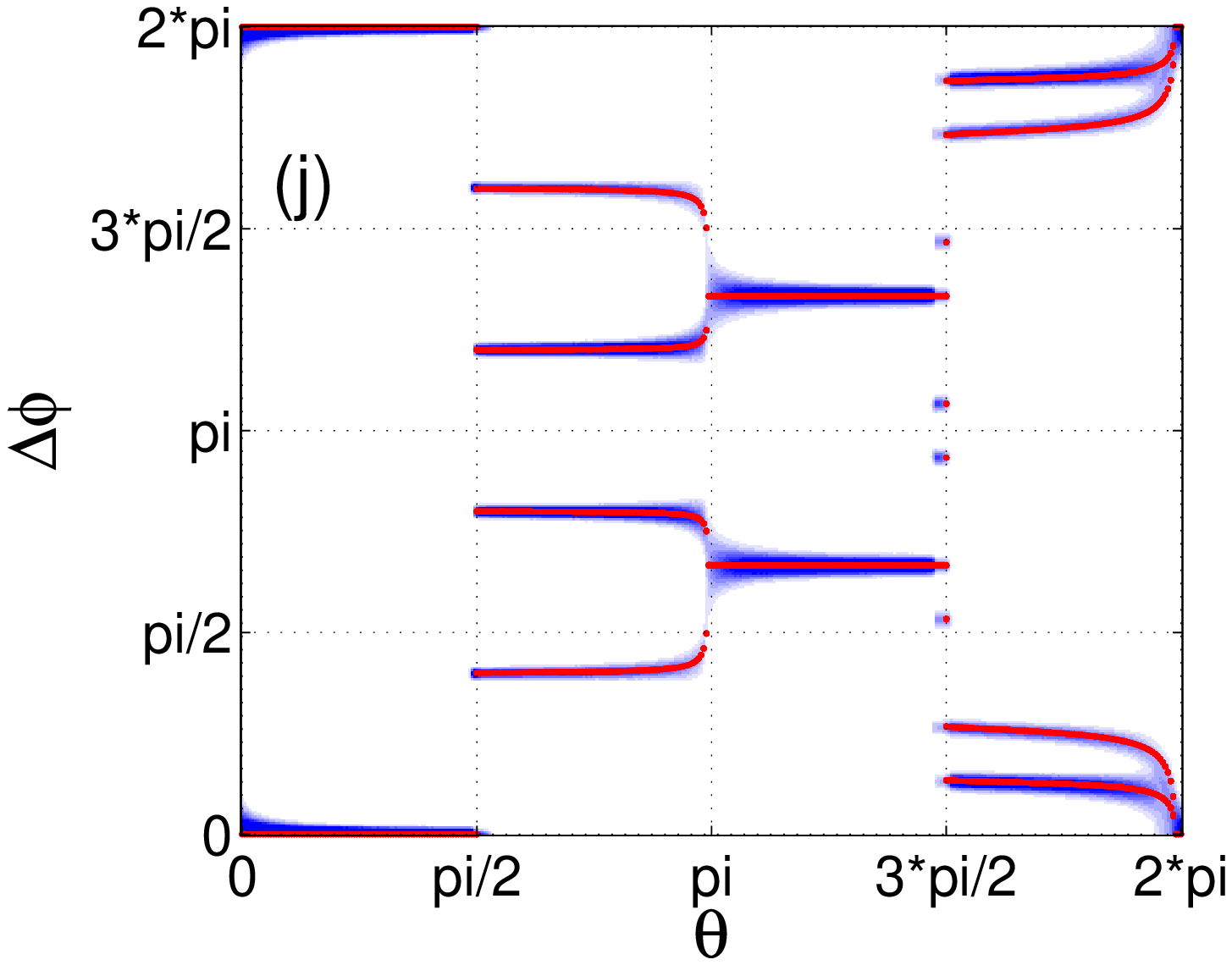}\label{fig:gs_hs_J1_J2_q5_phi}}
\subfigure{\includegraphics[scale=0.29,clip]{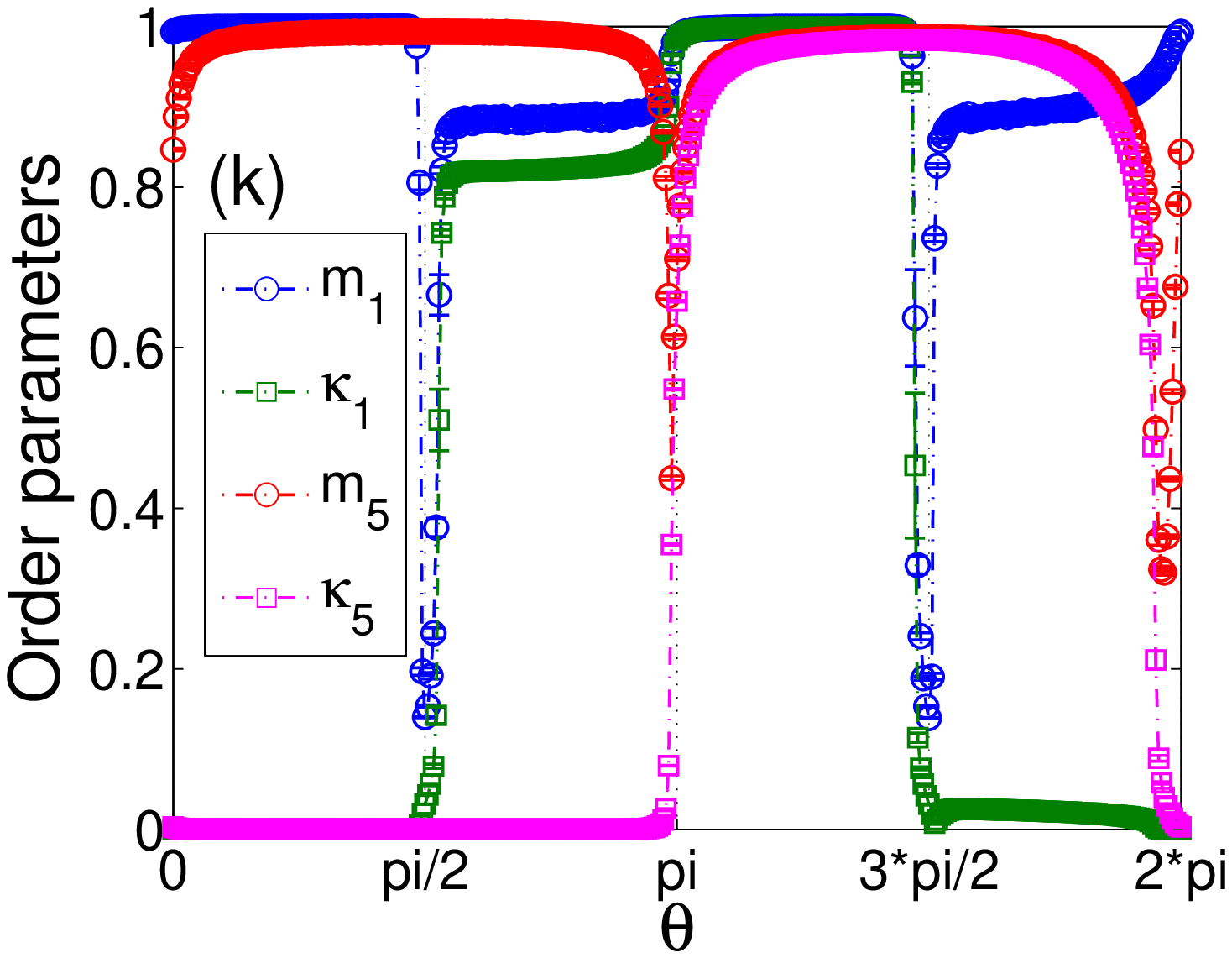}\label{fig:gs_J1_J2_q5_op}}
\subfigure{\includegraphics[scale=0.29,clip]{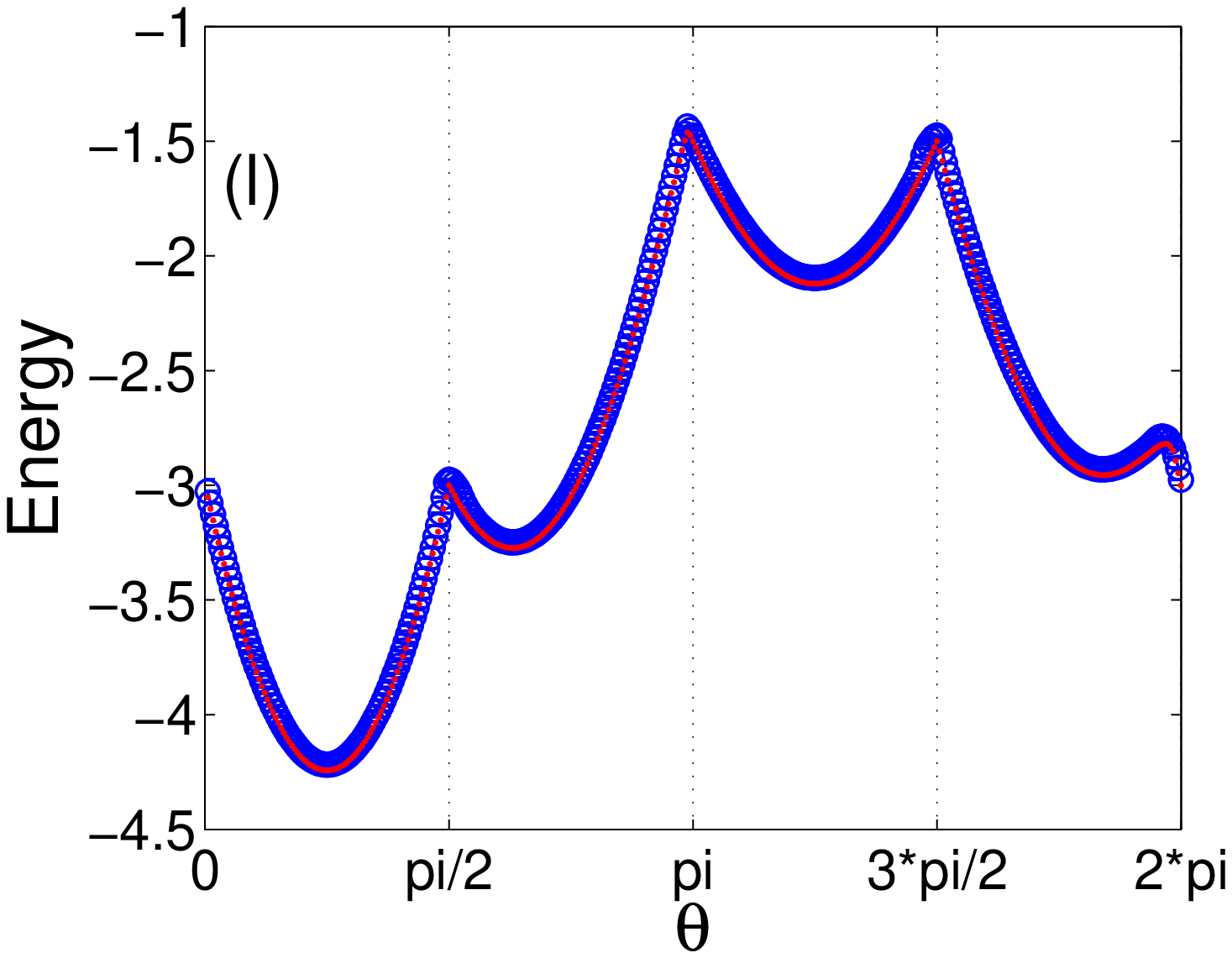}\label{fig:gs_J1_J2_q5_e}}\\ \vspace{-4mm}
\subfigure{\includegraphics[scale=0.29,clip]{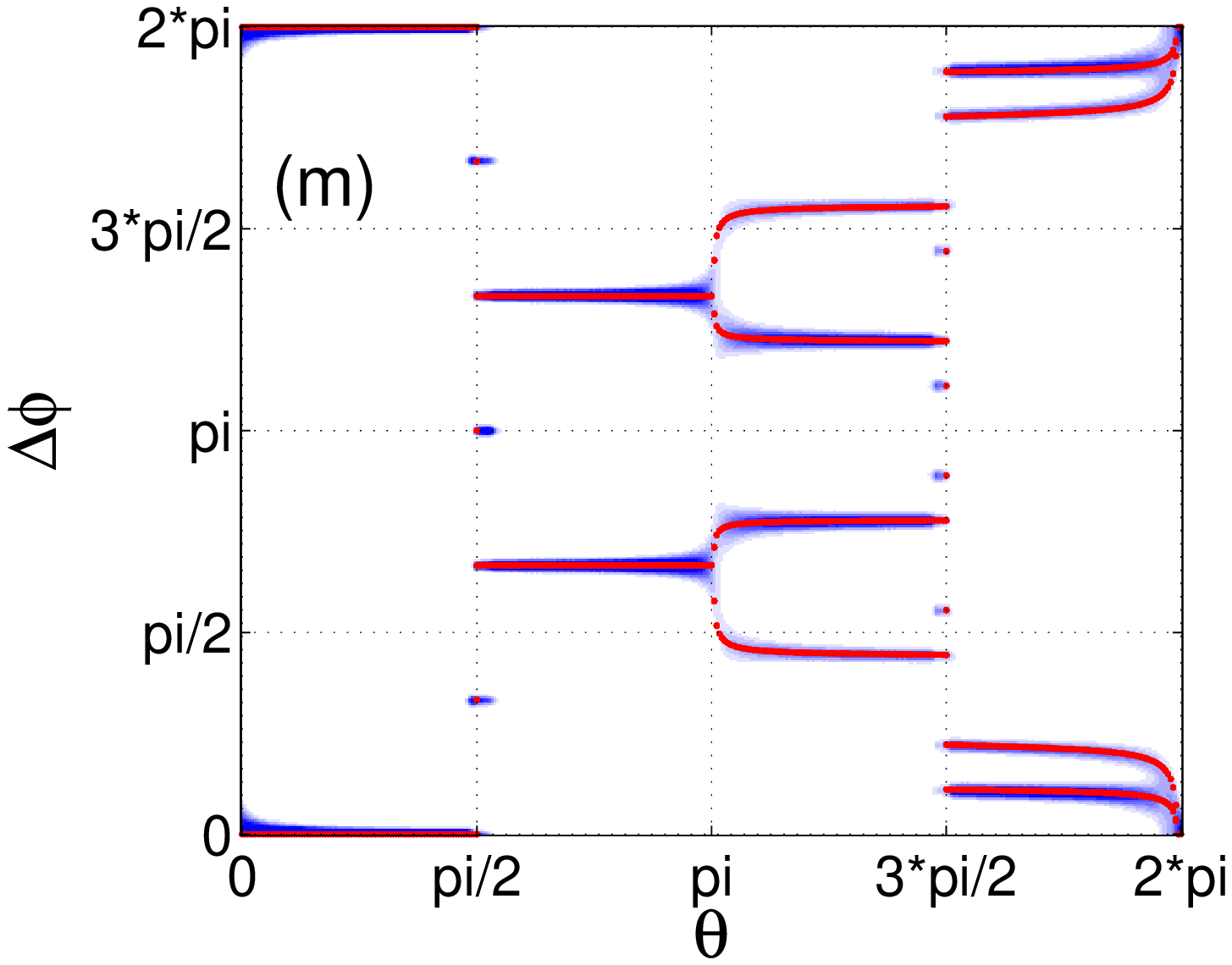}\label{fig:gs_hs_J1_J2_q6_phi}}
\subfigure{\includegraphics[scale=0.29,clip]{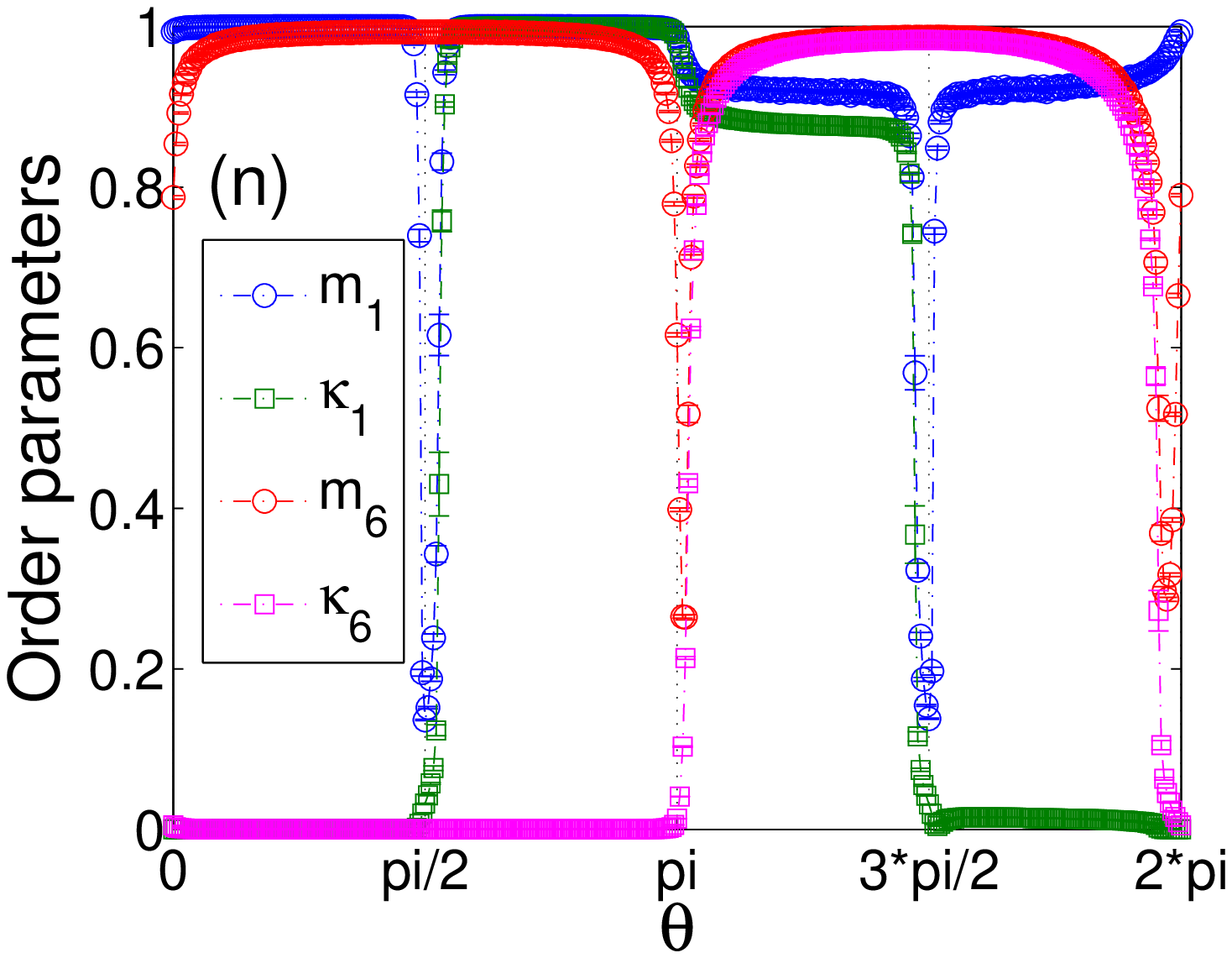}\label{fig:gs_J1_J2_q6_op}}
\subfigure{\includegraphics[scale=0.29,clip]{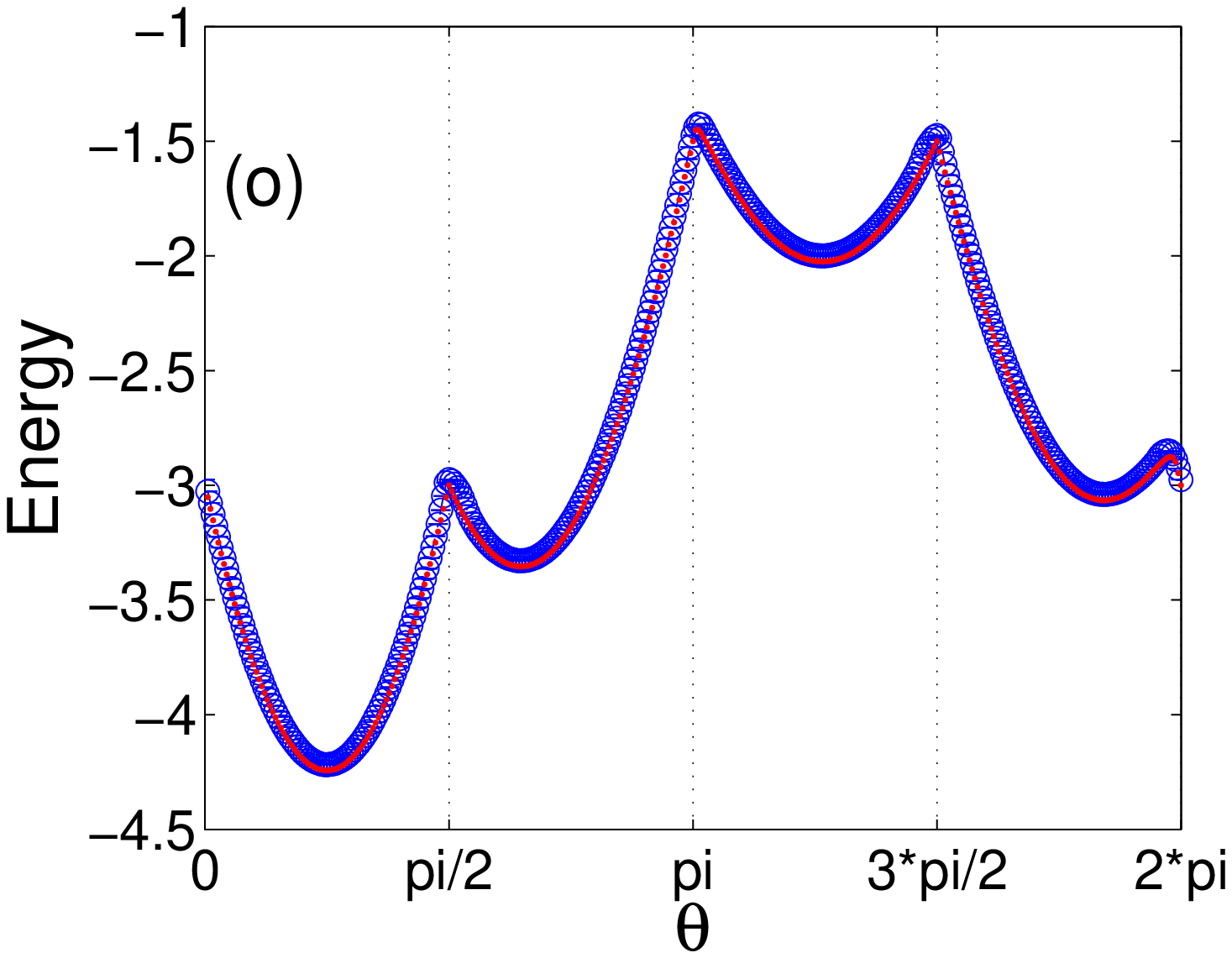}\label{fig:gs_J1_J2_q6_e}}\\ \vspace{-4mm}
\caption{(Color online) The relative phase angle $\Delta \phi$, the order parameters $m_1$, $m_q$, $\kappa_1$, $\kappa_q$, and the energy $e$, as functions of the exchange interaction ratio $J_q/J_1=\tan(\theta)$, for (a)-(c) $q=2$, (d)-(f) $q=3$, (g)-(i) $q=4$, (j)-(l) $q=5$, and (m)-(o) $q=6$. In (a), (d), (g), (j), and (m), the red bands represent top-view histograms (densities) obtained from MC simulations and the superimposed black dots show the Hamiltonian optimization values. In (c), (f), (i), (l), and (o), the blue circles represent the MC and the red dots the optimization results.}\label{fig:models}
\end{figure}

Below we present results covering the entire interaction parameters space $J_1,J_q\in[-1,1]$, corresponding to the variation of $\theta \in [0,2\pi)$. In particular, in Fig.~\ref{fig:models} we show the phase angle $\Delta \phi$, the order parameters $m_1$, $m_q$, $\kappa_1$, $\kappa_q$, and the energy $e$, as functions of $\theta$, for $q=2,\ldots, 6$. The phase angles (left column) in the form of histograms obtained from MC simulations (areas in blue) are complemented with the Hamiltonian minimization results (red curves). The order parameters (central column) are calculated from MC simulations and the equilibrium conditions are confirmed by comparing the internal energies (right column) from MC simulations (thick blue curve) with the residual values of the Hamiltonian minimization (thin red curve), which are found to practically collapse on the same curve.   

\begin{table*}
	\centering
		\begin{tabular}{l|cccc}
			      $q$   & $[\theta_{I,min},\theta_{I,max}]$  & $[\theta_{II,min},\theta_{II,max}]$ & $[\theta_{III,min},\theta_{III,max}]$ & $[\theta_{IV,min},\theta_{IV,max}]$ \\ \hline
			$2$    & [-14,90]  & [90,173] & [174,270] & [270,345] \\
			$3$    & [-6,90]  & [90,183] & [184,270] & [270,353] \\
			$4$    & [-3,90]  & [90,176] & [177,270] & [270,356] \\
			$5$    & [-2,90]  & [90,178] & [179,270] & [270,357] \\
			$6$    & [-1,90]  & [90,180] & [180,270] & [270,358] \\
			$7$    & [-1,90]  & [90,180] & [180,270] & [270,358] \\
			$8$    & [0,90]  & [90,180] & [180,270] & [270,359]
		\end{tabular}
\caption{Ranges of $\theta$ (in degrees) defining four different phases in the parameter plane $J_1-J_q$, for $q=2,\ldots,8$.}\label{intervals}
\end{table*}

For each $q$, four phases (intervals of $\theta$), corresponding to different types of ordering, can be distinguished. The extents of the respective phases slightly change with $q$ and are summarized in Table~\ref{intervals}, for $q=2,\ldots, 8$. We note that the boundary values for $q=2$ are consistent~\footnote{One should take into account slightly different Hamiltonians here and in Ref.~\cite{zuko03b}.} with those determined for the in-plane angles in the three-dimensional model~\cite{zuko03b}. 

The first phase of the FM quasi-LRO corresponding to the interval $[\theta_{I,min},\theta_{I,max}]$ covers the entire quadrant of $J_1>0$, $J_q>0$. It also partially spreads to the quadrant of $J_1>0$, $J_q<0$ but its extent is gradually diminished with increasing $q$ and eventually vanishes at $q=8$. In the present calculations we could not see any significant differences between these FM phases for different values of $q$. However, we note that the finite temperature calculations pointed to the change of the phase diagram topology, giving different kinds of FM ordering for $q<5$ and for $q \ge 5$~\cite{pode11}.

\begin{figure}[t!]
\centering
\subfigure{\includegraphics[scale=0.44,clip]{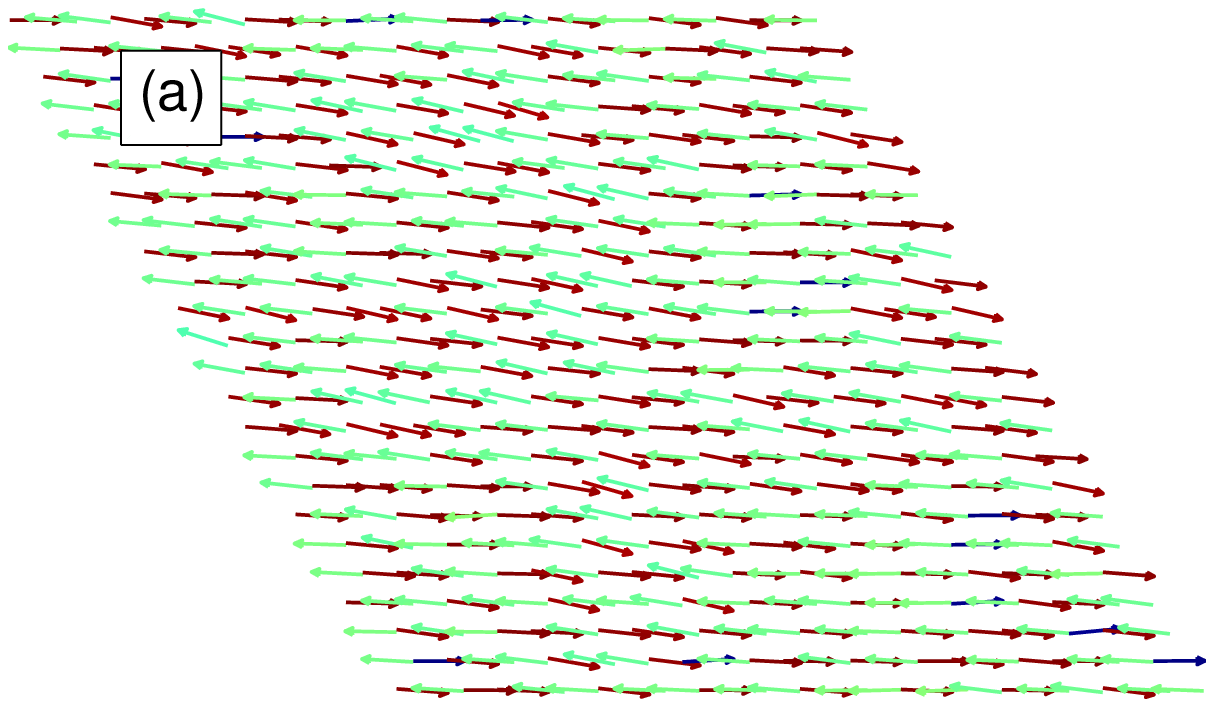}\label{fig:snap_q2_II}}
\subfigure{\includegraphics[scale=0.44,clip]{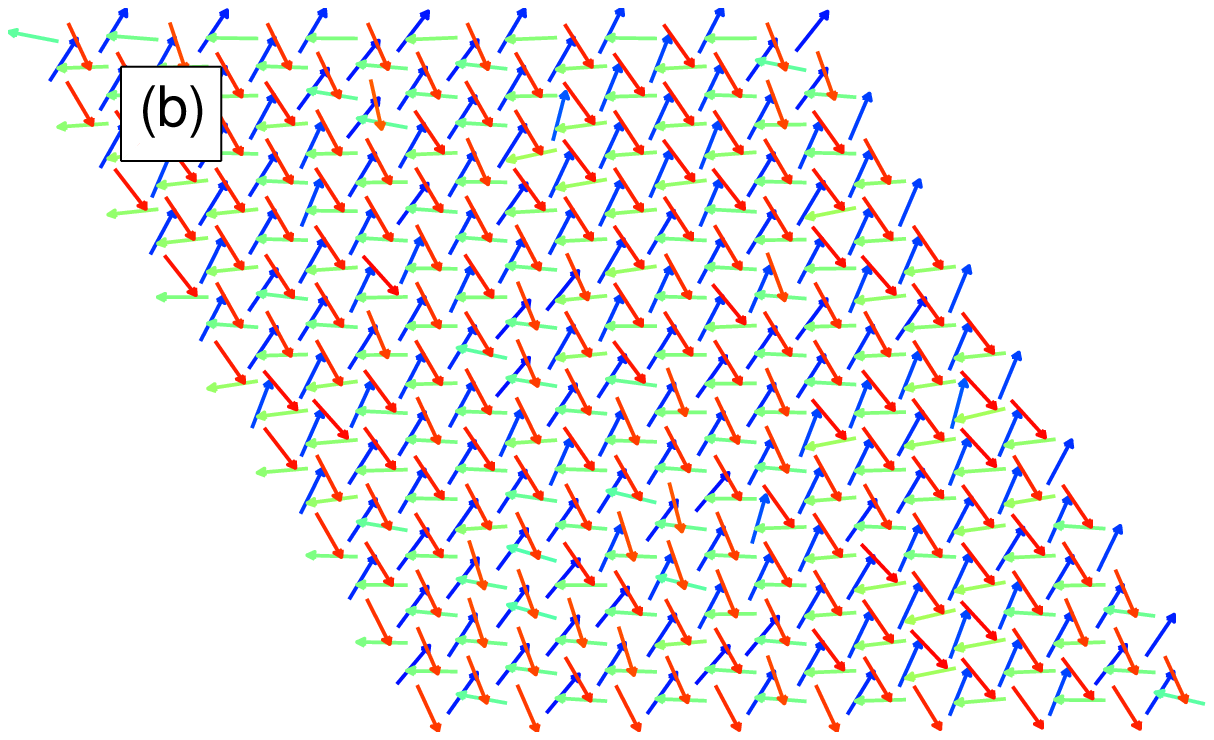}\label{fig:snap_q2_III}}
\subfigure{\includegraphics[scale=0.44,clip]{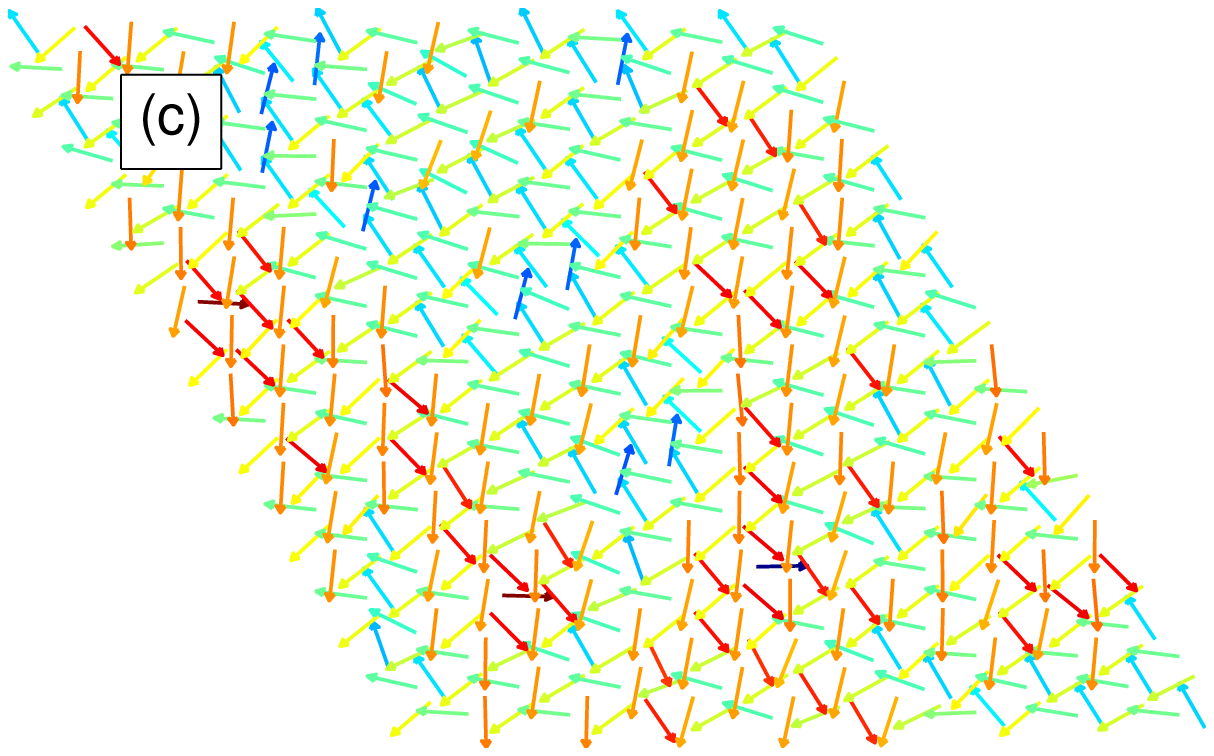}\label{fig:snap_q2_IV}}\\
\subfigure{\includegraphics[scale=0.44,clip]{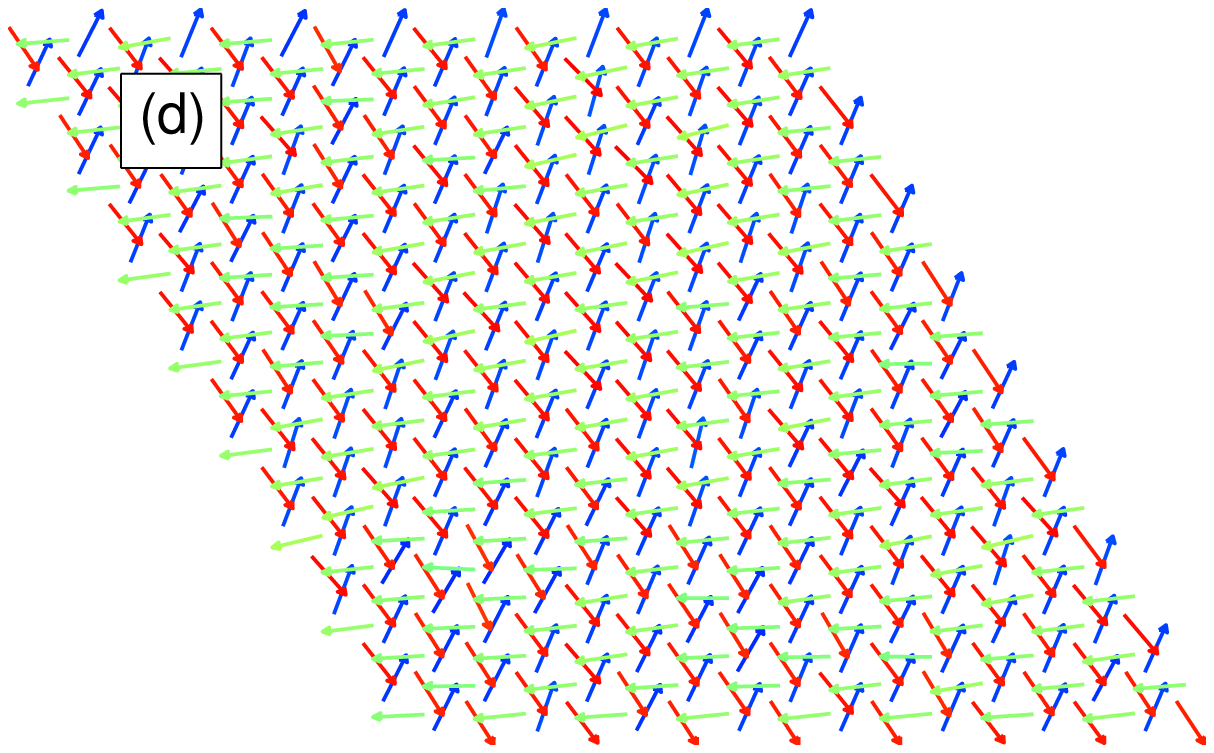}\label{fig:snap_q3_II}}
\subfigure{\includegraphics[scale=0.44,clip]{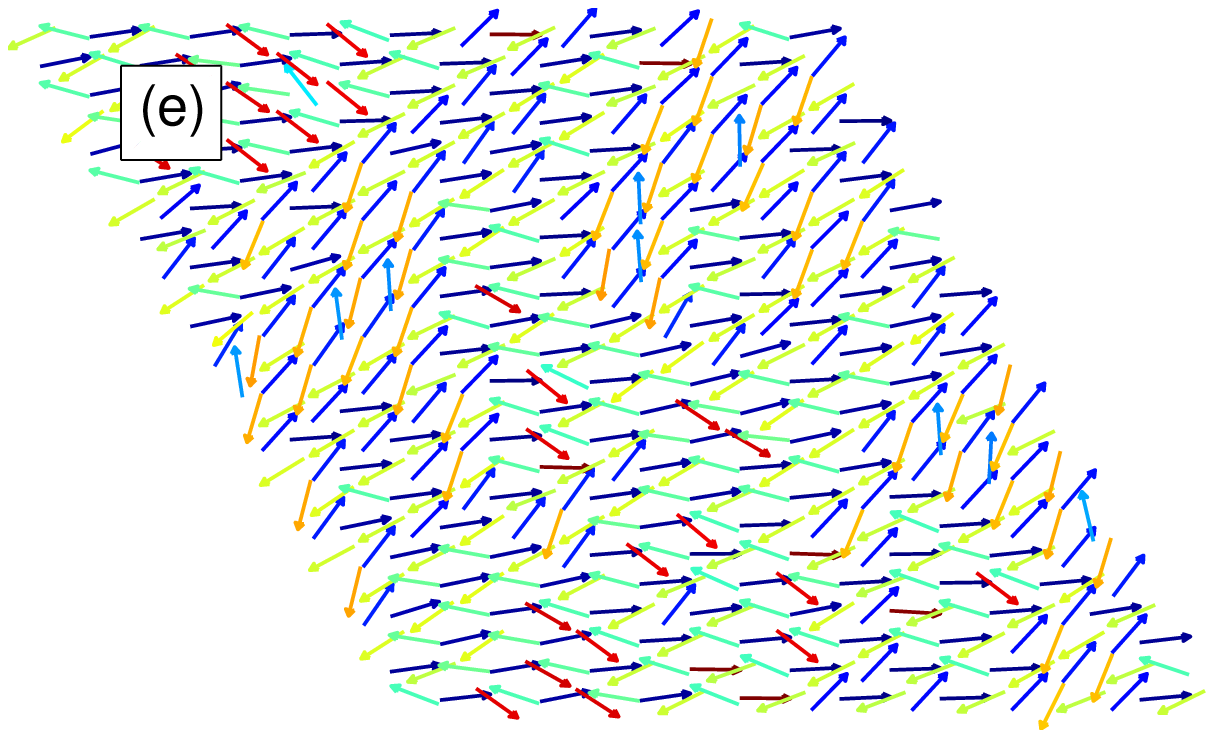}\label{fig:snap_q3_III}}
\subfigure{\includegraphics[scale=0.44,clip]{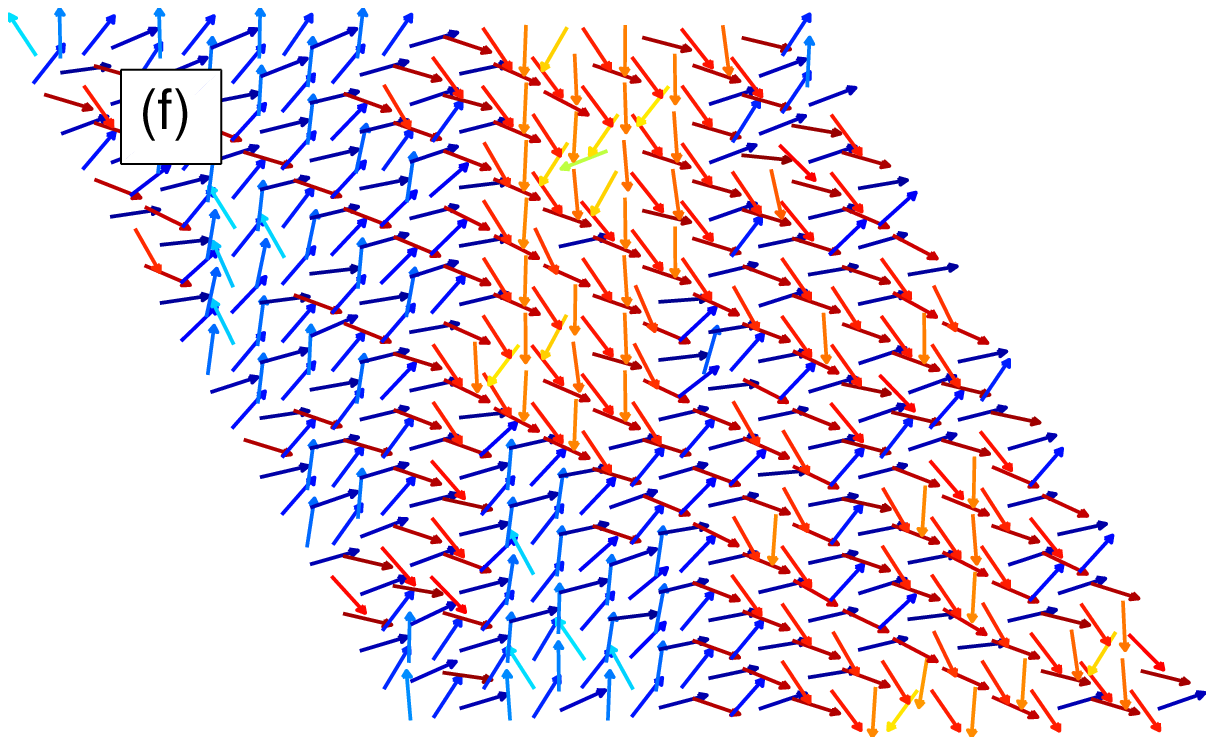}\label{fig:snap_q3_IV}}\\
\subfigure{\includegraphics[scale=0.44,clip]{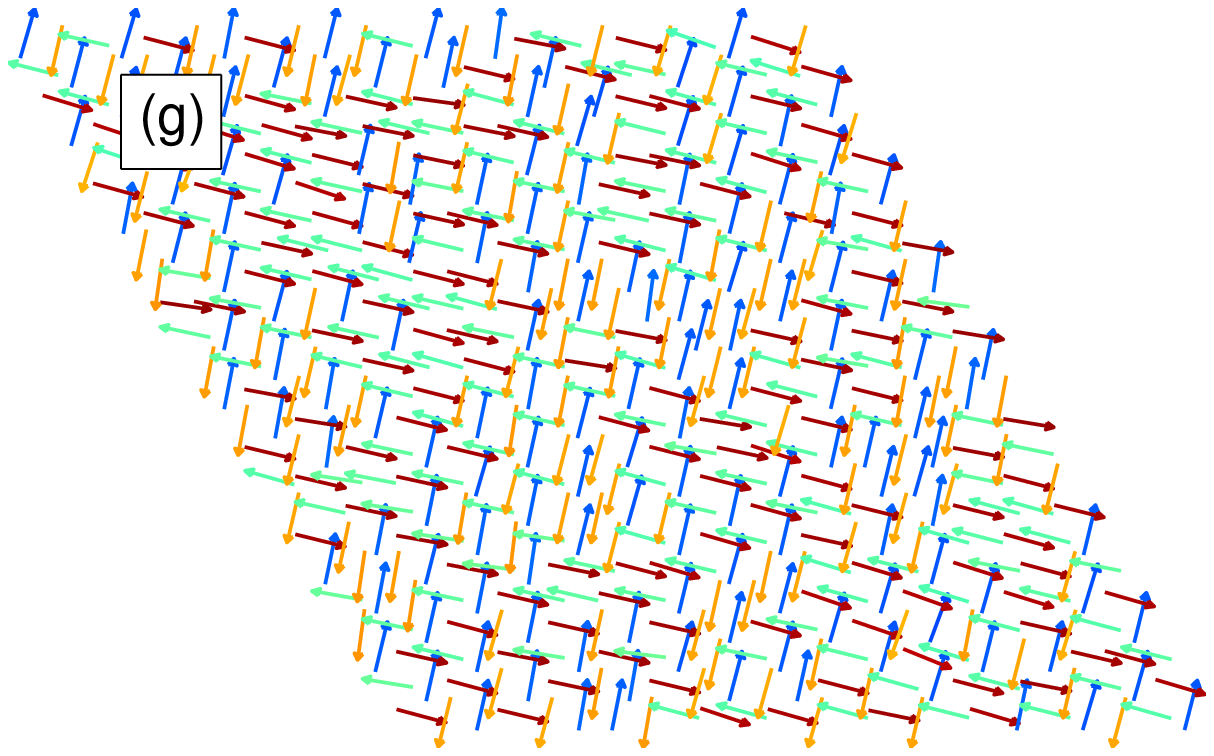}\label{fig:snap_q4_II}}
\subfigure{\includegraphics[scale=0.44,clip]{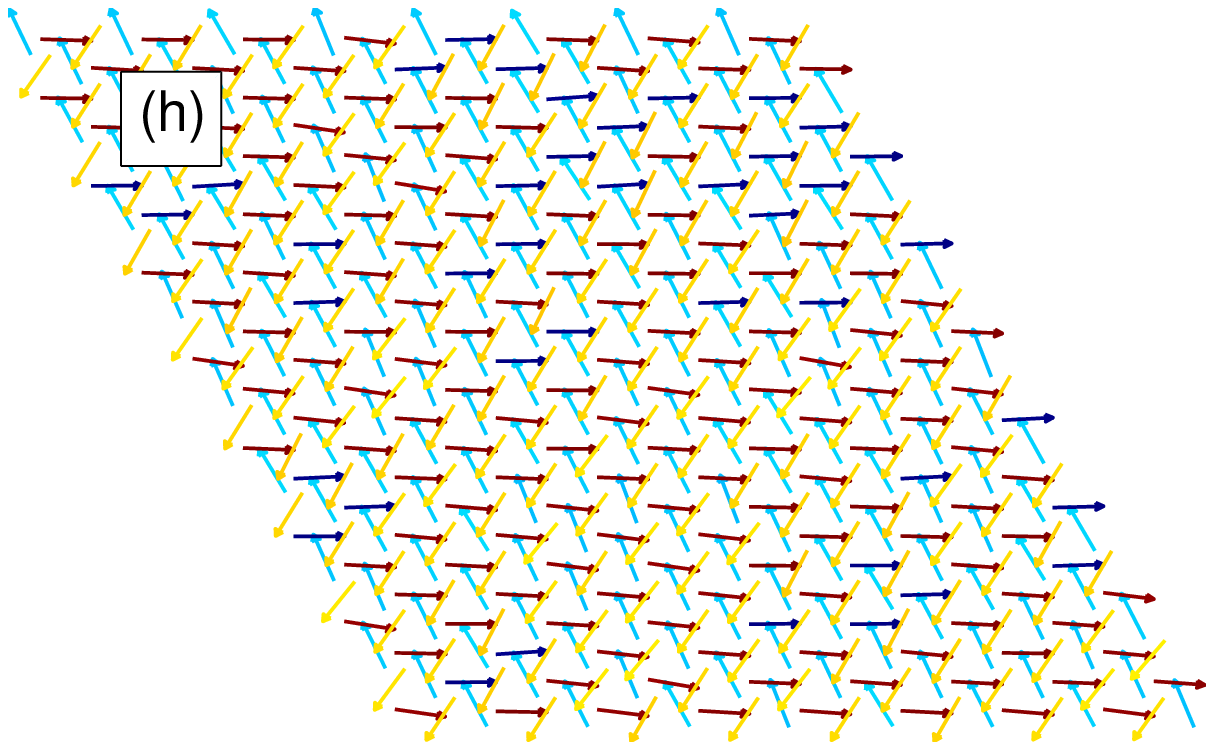}\label{fig:snap_q4_III}}
\subfigure{\includegraphics[scale=0.44,clip]{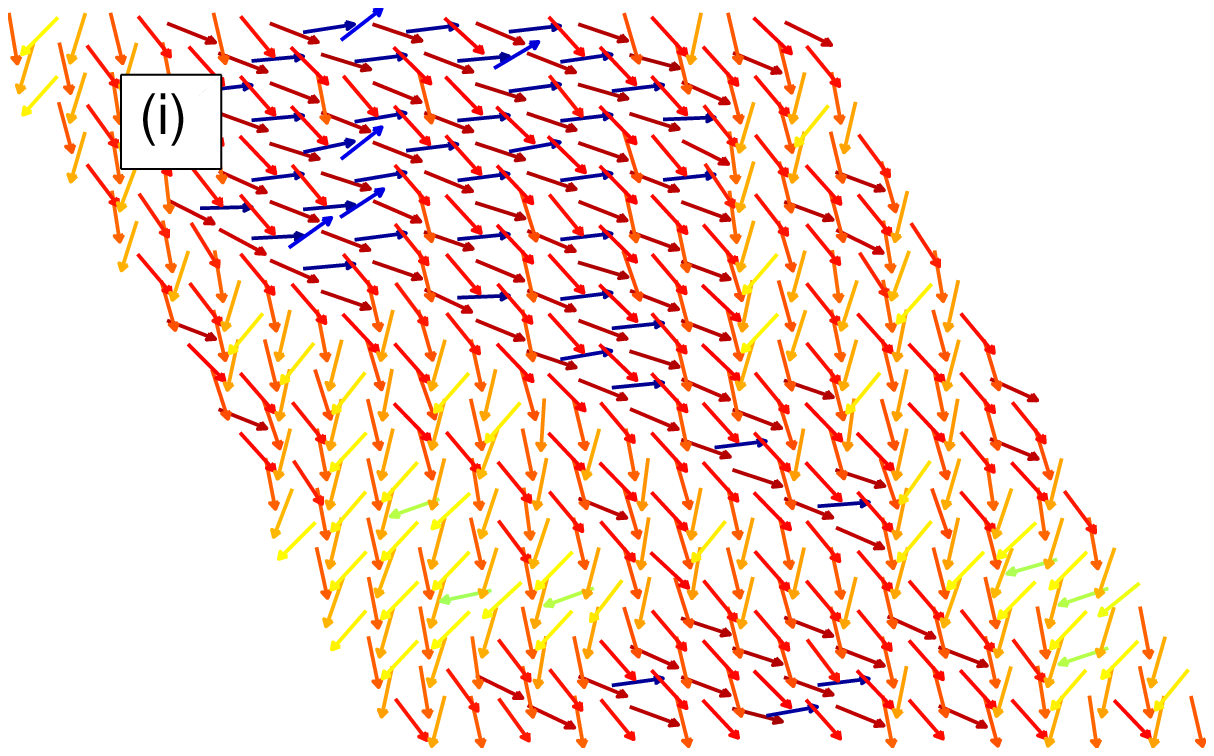}\label{fig:snap_q4_IV}}\\
\subfigure{\includegraphics[scale=0.44,clip]{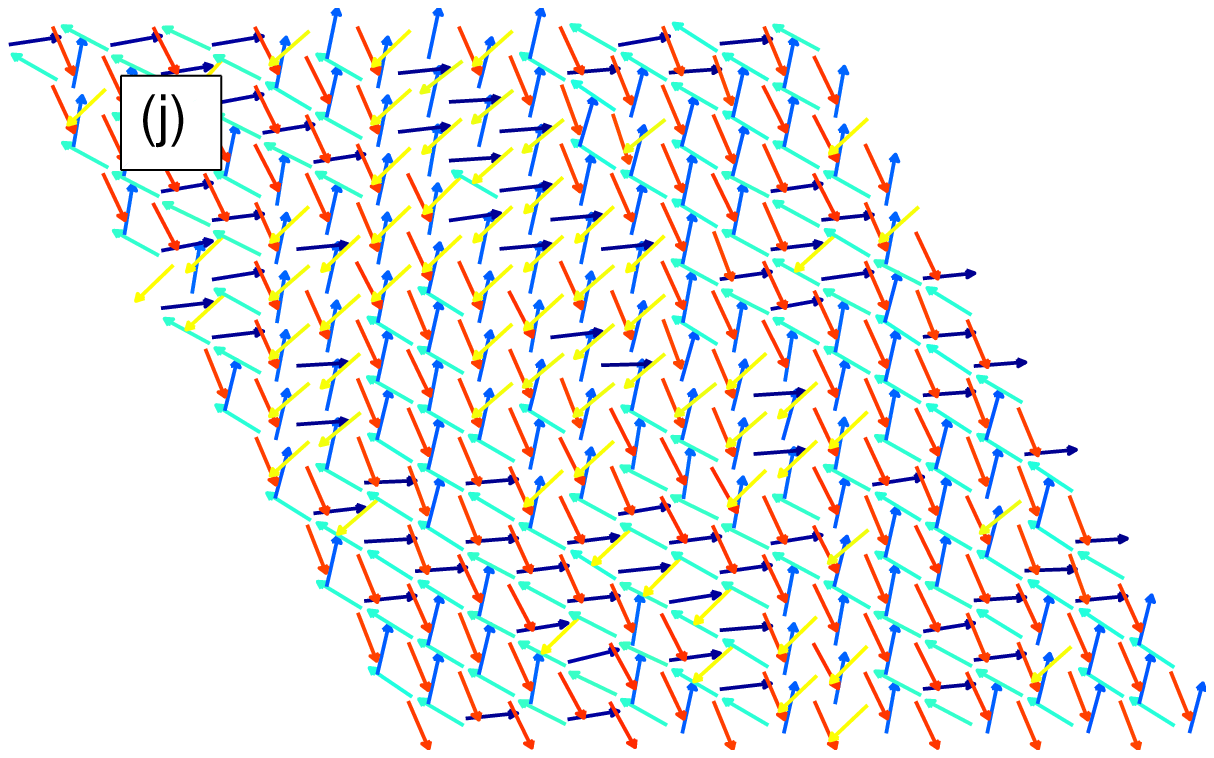}\label{fig:snap_q5_II}}
\subfigure{\includegraphics[scale=0.44,clip]{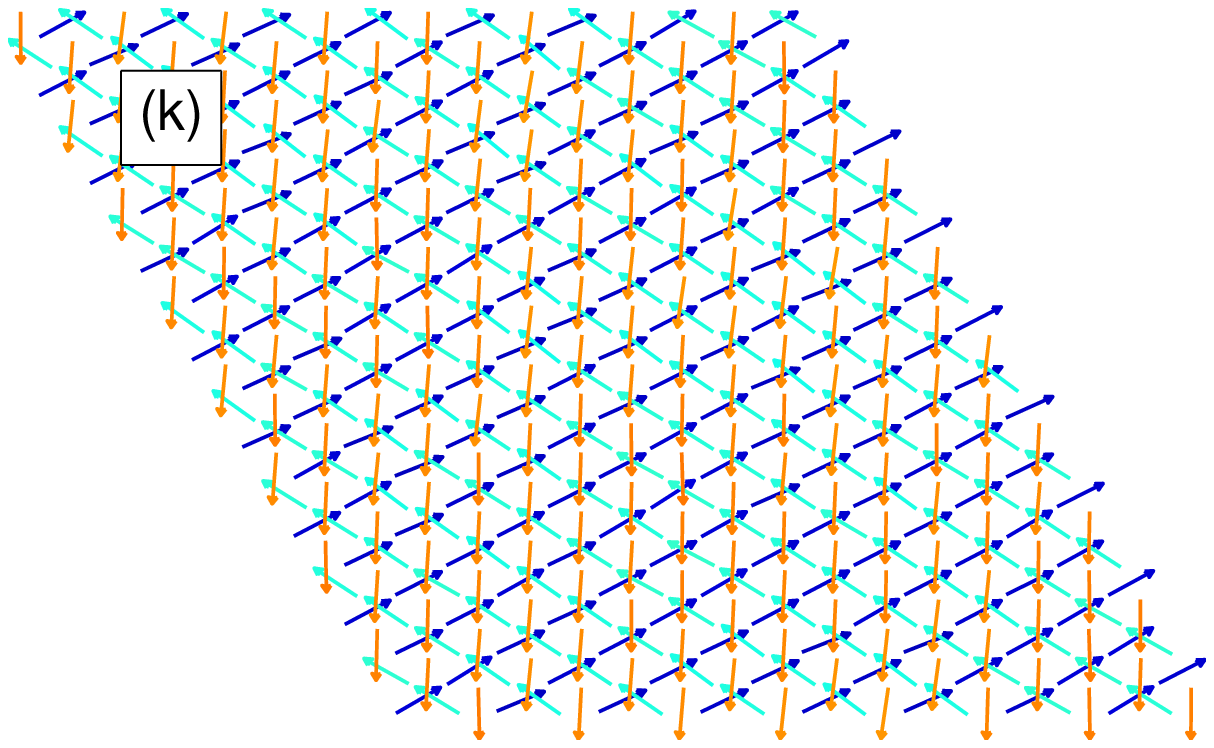}\label{fig:snap_q5_III}}
\subfigure{\includegraphics[scale=0.44,clip]{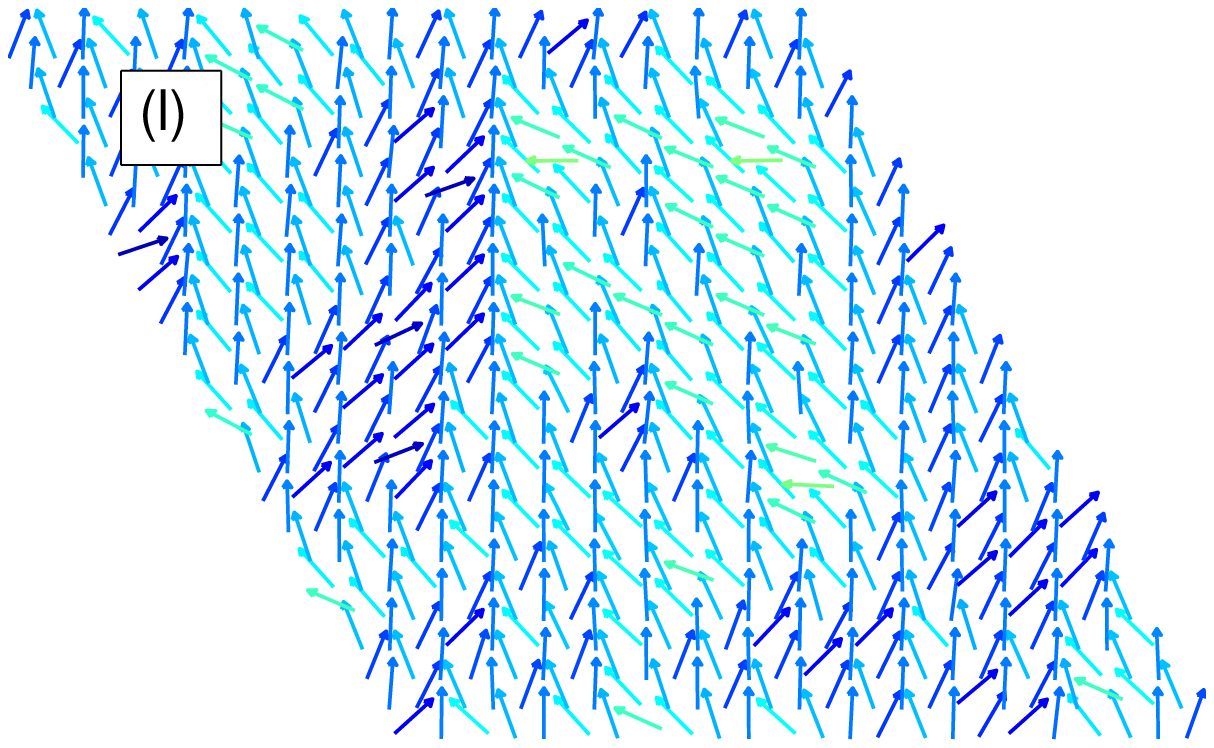}\label{fig:snap_q5_IV}}\\
\subfigure{\includegraphics[scale=0.44,clip]{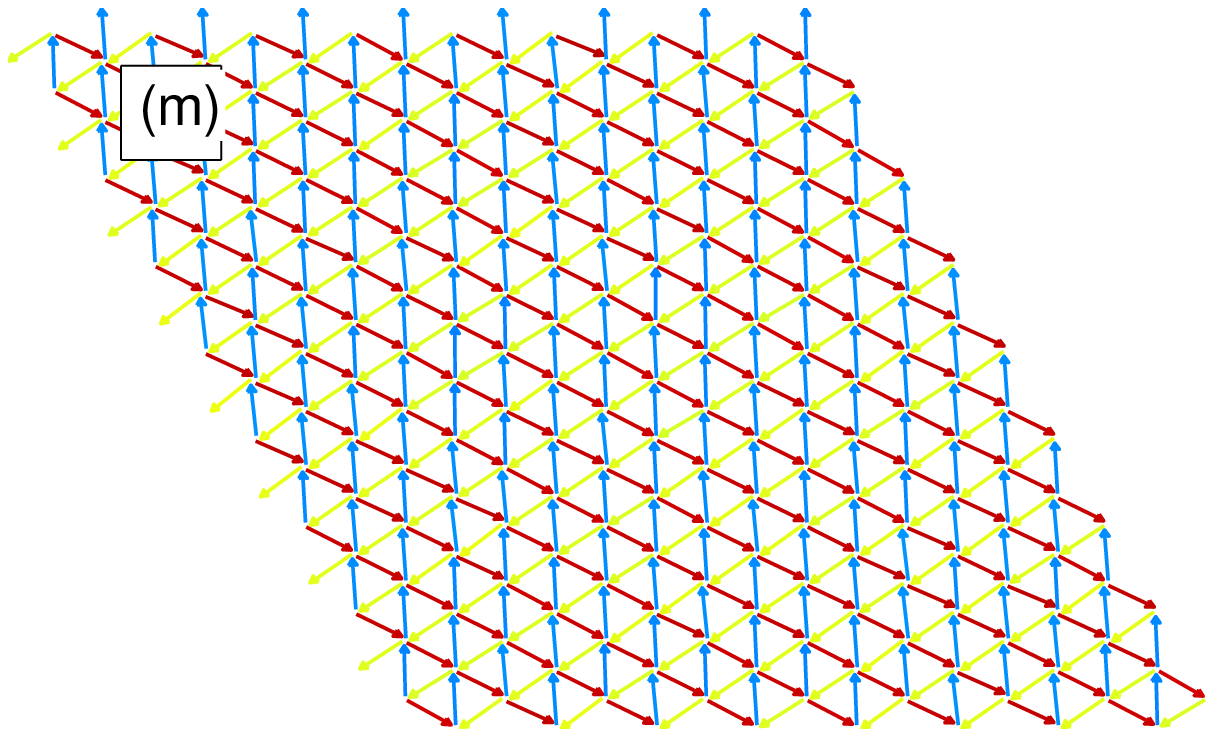}\label{fig:snap_q6_II}}
\subfigure{\includegraphics[scale=0.44,clip]{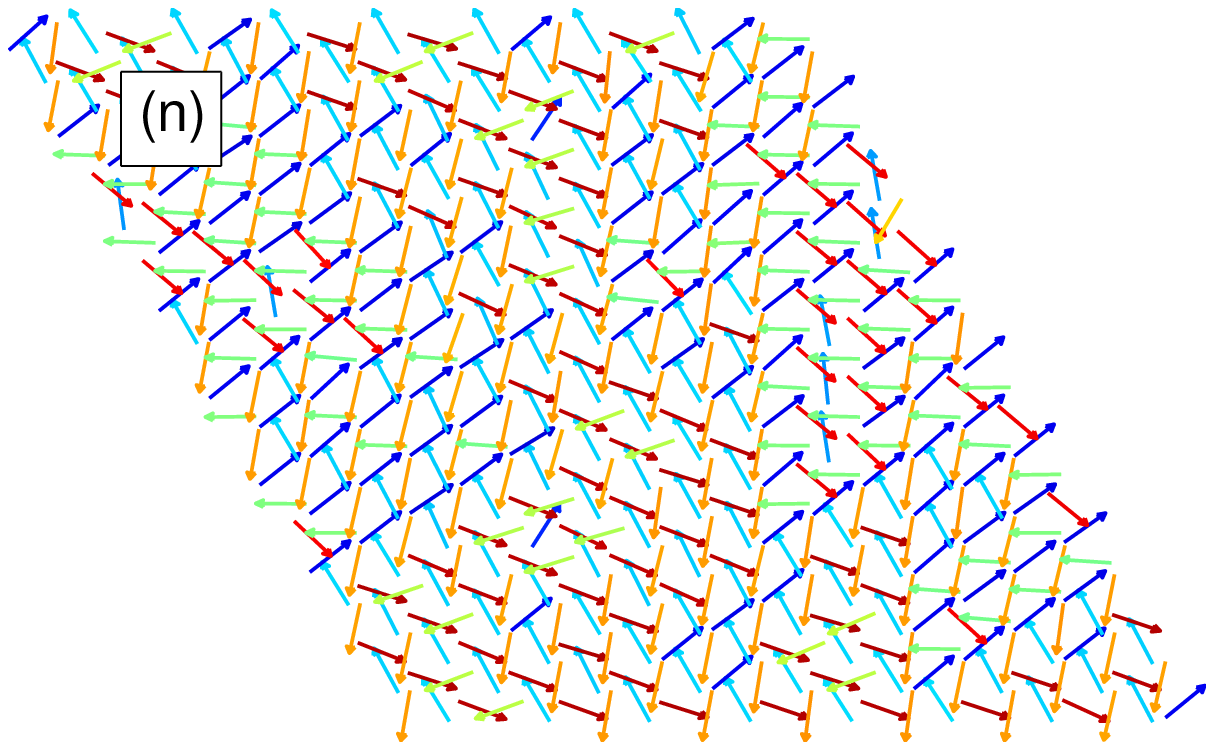}\label{fig:snap_q6_III}}
\subfigure{\includegraphics[scale=0.44,clip]{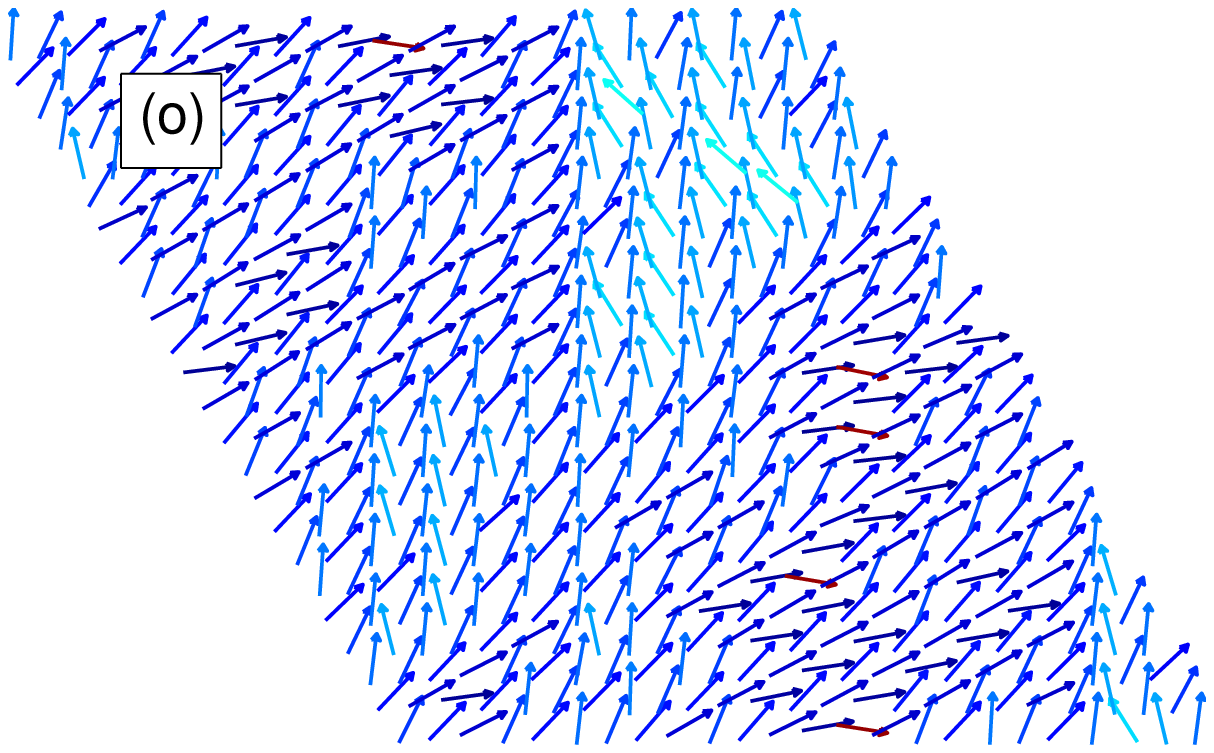}\label{fig:snap_q6_IV}}
\caption{(Color online) Near-ground-state ($T=0.05$) spin snapshots in the frustrated phases taken at $\theta_{II}=3\pi/4$ (left column), $\theta_{III}=5\pi/4$ (central column) and $\theta_{IV}=7\pi/4$ (right column), for (a)-(c) $q=2$, (d)-(f) $q=3$, (g)-(i) $q=4$, (j)-(l) $q=5$, and (m)-(o) $q=6$.}\label{fig:snaps}
\end{figure}

\begin{figure}[t!]
\centering
\subfigure{\includegraphics[scale=0.4]{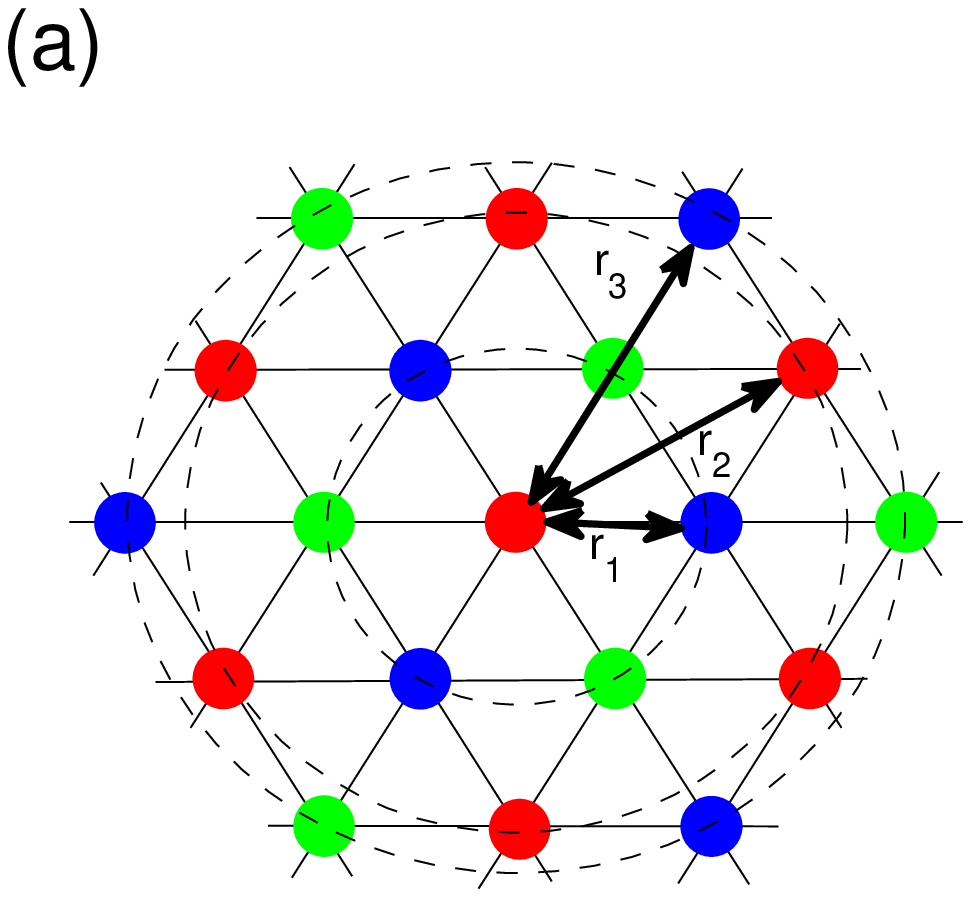}\label{fig:latt}}
\subfigure{\includegraphics[scale=0.4,clip]{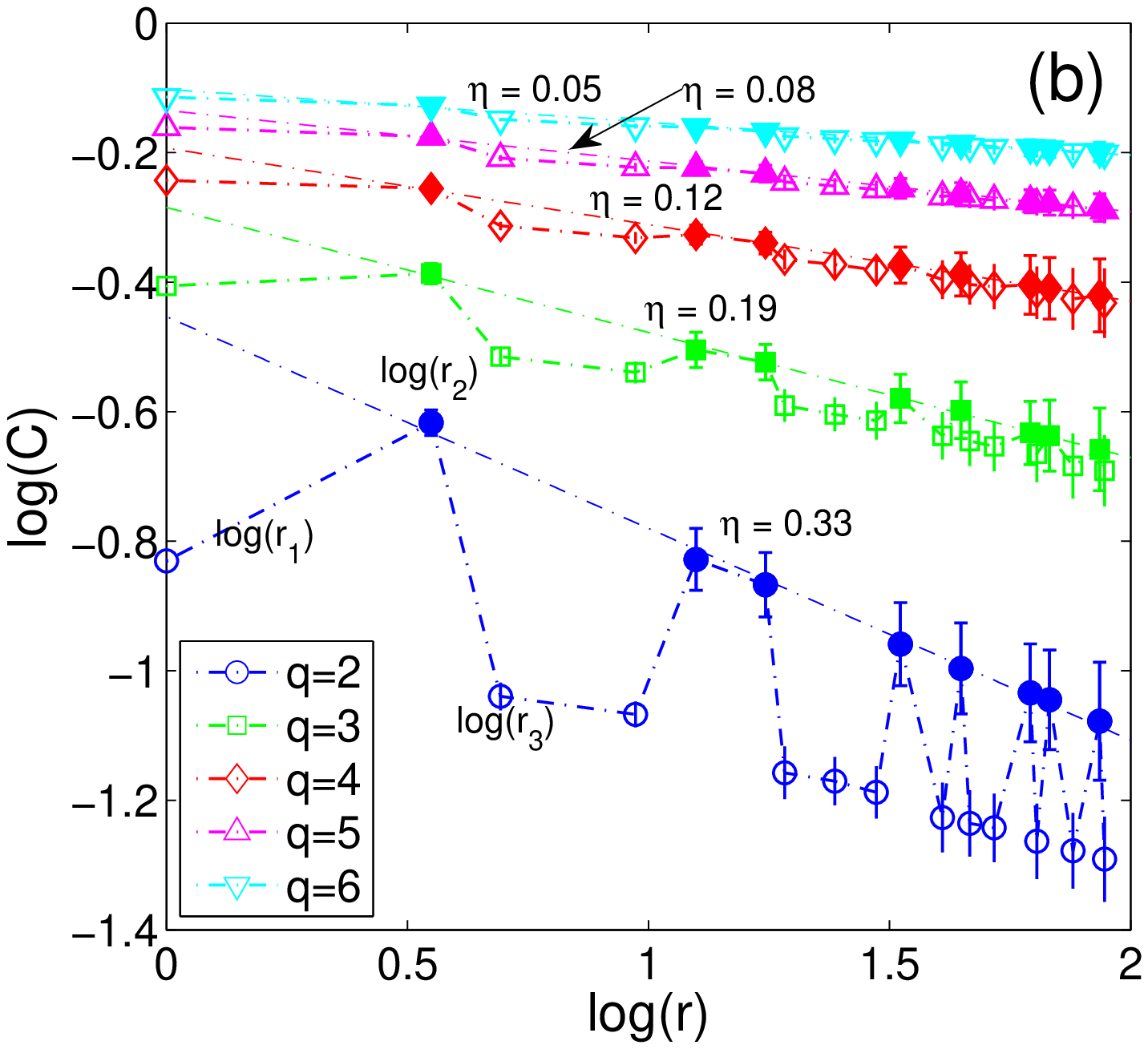}\label{cf1_7pi_4}}\\
\subfigure{\includegraphics[scale=0.4,clip]{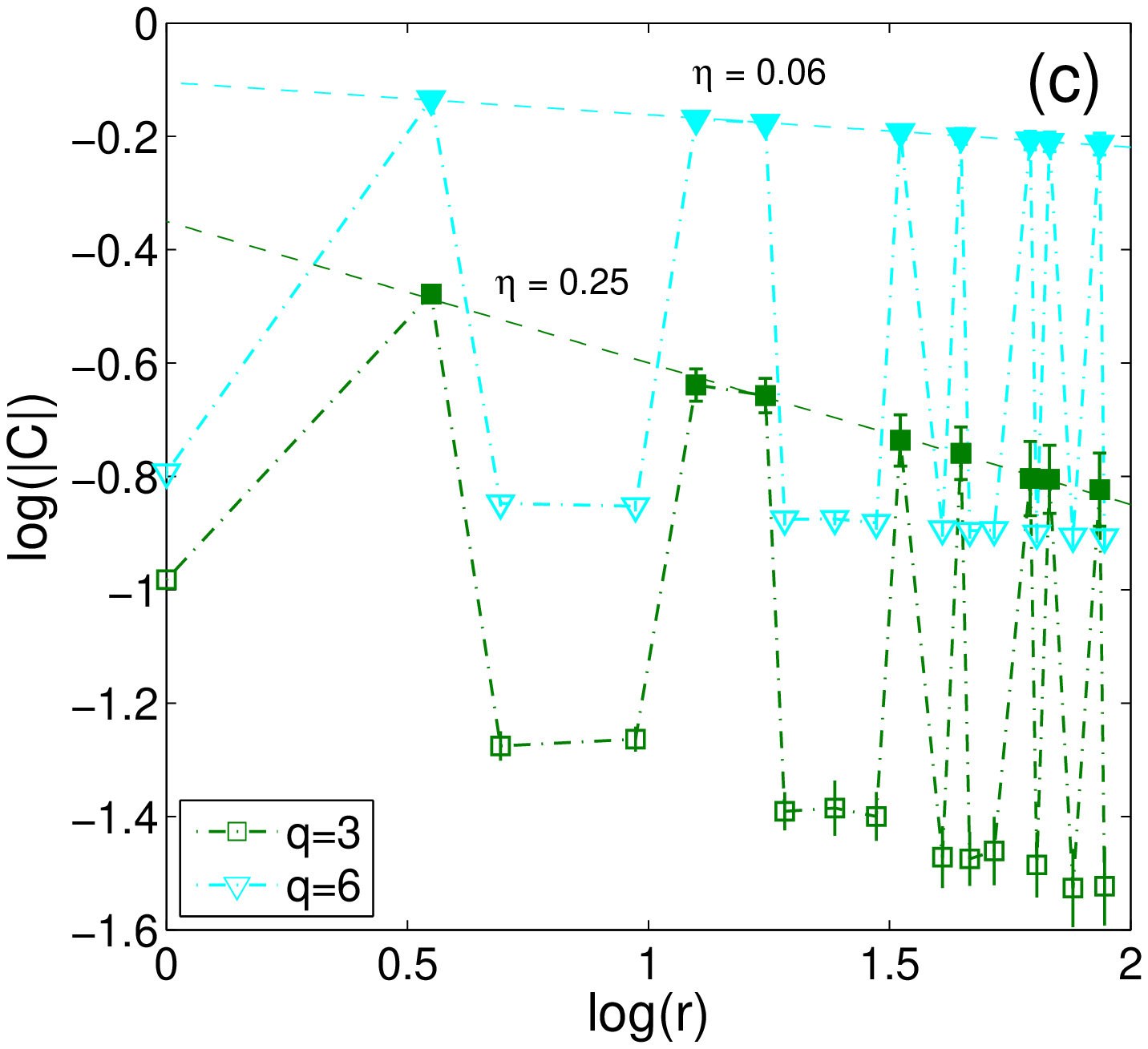}\label{cf1_5pi_4}}
\subfigure{\includegraphics[scale=0.4,clip]{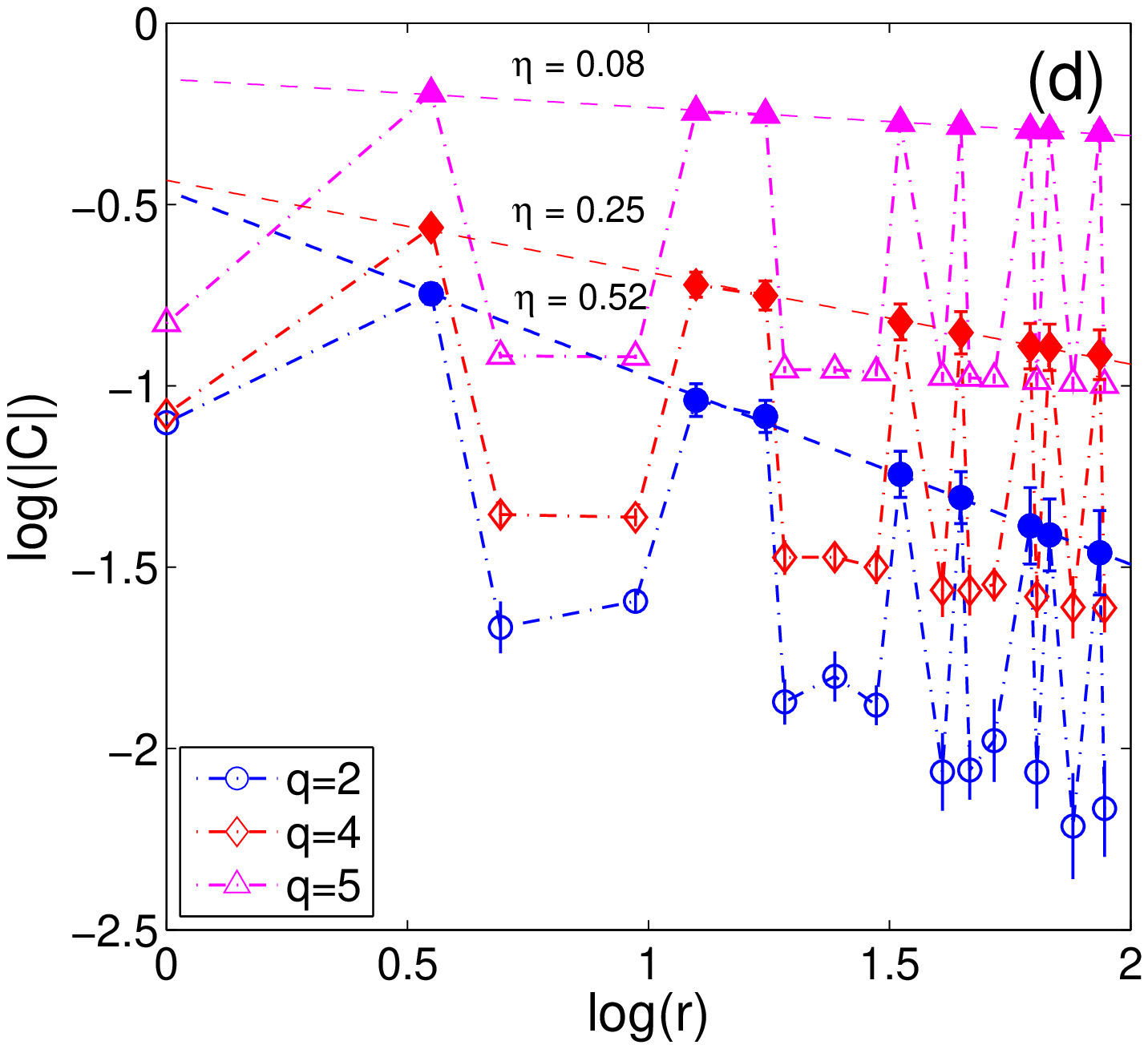}\label{cf1_3pi_4}}
\caption{(Color online) (a) Three different color-coded sublattices of the triangular lattice with the distances $r_1,r_2$ and $r_3$ between the first-, second- and third-nearest neighbors, respectively. Spin correlation functions for $q=2,\ldots,6$, $L=48$ and $T=0.05$ on a log-log scale, for (b) $\theta_{IV}=7\pi/4$, (c) $\theta_{III}=5\pi/4$, and (d) $\theta_{II}=3\pi/4$. In (c),(d) the empty symbols correspond to negative values of $C$. The values on the lags corresponding to the pairs of spins belonging to the same sublattice are denoted by the filled symbols and are fitted to the power law function with the exponent $\eta$.}\label{fig:cf}
\end{figure}

The situation in the remaining intervals is more complex due to the presence of geometrical frustration and/or competition between the interactions $J_1$ and $J_q$. To better illustrate the nature of the ordering in those intervals we additionally present in Fig.~\ref{fig:snaps} spin snapshots and in Fig.~\ref{fig:cf} the corresponding spin pair correlation functions. Let us continue analyzing the situation when $J_1$ remains ferromagnetic but $J_q$ is changed to negative values, i.e. the interval IV. As evidenced from Fig.~\ref{fig:models}, for $\theta \in [\theta_{IV,min},\theta_{IV,max}]$, the ferromagnetic phase angle $\Delta \phi=0$ splits at $\theta_{IV,max}$ to some $\theta$-dependent values $\pm\Delta \phi_1(\theta),\pm\Delta \phi_2(\theta)$\footnote{Since we consider $\Delta \phi \in [0,2\pi]$, due to the rotational symmetry $-\Delta \phi \to 2\pi-\Delta \phi$}, which tend to the phase angles $\Delta\phi(\theta_{IV,min})\in\{2i\pi/3q,i=1,\ldots,3q\}\setminus\{2j\pi/q,j=1,\ldots,q\}$ of the purely ($\theta_{IV,min}=3\pi/2$) generalized nematic phase, for each value of $q$. Note that as long as the ferromagnetic interactions are non-zero, the allowed phase angles $\pm\Delta \phi_1(\theta),\pm\Delta \phi_2(\theta)$ are reduced to those with the  smallest absolute values, thus enabling the best possible collinear alignment between the nearest neighbors, while the remaining possible generalized nematic states are suppressed. This phase is characterized by finite but unsaturated values of the order parameters, except for $m_q$ and $\kappa_q$ that saturate in the limiting value of $\theta_{IV,min}$ (Fig.~\ref{fig:models}). 

The corresponding snapshots in the right column of Fig.~\ref{fig:snaps} indicate a certain degree of FM ordering that, however, does not spread over the entire lattice but is rather contained within smaller domains. Nevertheless, with the increasing $q$ the domain sizes tend to increase, which is also reflected in the increasing values of the magnetic order parameter $m_1$. This can be explain in terms of a gradual relaxation of the frustration between the conflicting interactions $J_1>0,J_q<0$ due to the increase of the degrees of freedom of the generalized nematic order. Therefore, in the limit of $q\to \infty$ the phase angle $\Delta \phi \to 0$ and one can expect a full recovery of the standard FM quasi-LRO. The increasing ferromagnetic correlations with the increasing $q$ are also evident from the behavior of the spin correlation function, presented in Fig.~\ref{cf1_7pi_4}. One can also notice that the presence of the generalized nematic term with $J_q<0$ decreases the correlation at the lags corresponding to distances between different sublattices (empty symbols), such as at the lags $r_1$ and $r_3$ shown in Fig.~\ref{fig:latt}. Nevertheless, the nature of the correlations remains algebraic, as for the standard XY ferromagnet, albeit, with increased values of the exponent $\eta(q,T)$.

Let us now focus on the case of both negative interactions or, more precisely, the case of $\theta \in [\theta_{III,min},\theta_{III,max}]$. The case of $J_1<0,J_2<0$ has already been studied~\cite{zuko03b,park08} and, in line with the present results for $q=2$, the ground state has been confirmed to be chiral AFM, characterized by the phase angles $\Delta \phi =\pm 2\pi/3$, for any ratio of $J_2/J_1$. However, for $q=3$ the picture changes drastically. The chiral AFM order disappears and the neighboring spins align forming turn angles with $\theta$-dependent values $\pm\Delta \phi_1(\theta),\pm\Delta \phi_2(\theta)$. This is similar to the phase IV but since $J_1$ is now AFM the preferred phase angles are those with the smallest and the largest absolute values. Thus, the correlation between spins belonging to the same (different) sublattices is positive (negative) (Fig.~\ref{cf1_5pi_4}). Like in the phase IV, the sublattice correlation function follows the power law with $\eta(q,T)$ larger than for the standard XY antiferromagnet, but decreasing with $q$. However, this is only true in the instances when the turn angles between different sublattices, preferred by the generalized nematic interactions for a given $q$, do not include the chiral AFM phase angles $\Delta \phi =\pm 2\pi/3$, as it is for $q=3$ or $q=6$. In all the other instances, like for $q=2,4,5$, there is no conflict between the magnetic and generalized nematic interactions and the systems show the chiral AFM ordering (see the central column in Fig.~\ref{fig:snaps}).

Finally, similar arguments can be made in order to explain ordering in the phase corresponding to $\theta \in [\theta_{II,min},\theta_{II,max}]$ but the dependence on $q$ is somehow reversed. Namely, the chiral AFM order is preserved for any $q$ divisible by 3, like for $q=3,6$ in our results, in which cases the spin alignments dictated by the magnetic and generalized nematic interactions are not antagonistic (see the left column in Fig.~\ref{fig:snaps}). Otherwise, i.e., for $q$ non-divisible by 3, there is a complex quasi-LRO, like in the phase III for $q=3,6$, with the algebraic correlations and increased values of the exponent $\eta(q,T)$ (Fig.~\ref{cf1_3pi_4}).

\subsection{Mapping between magnetic states and mixed-metal cyanide structures}

In the following, we establish mapping between the ground states of the above generalized frustrated XY model on a triangular lattice and the structural chemistry of bimetalic cyanides. By finding analogy between the magnetic interactions in the spin model and the supramolecular interactions in the chemical compound, a mapping between the XY model with $q=2$ and the structural chemistry of the compound ${\rm Ag}_{1/2}{\rm Au}_{1/2}{\rm (CN)}$ was recently established by Cairns at al.~\cite{cairns16} Here, we extend this approach for the mapping between the model with $q>2$ and the cyanid compounds with appropriate mixing patterns of the metallic cations.

The pure AuCN and AgCN structures consist of the linear chains with the strictly repeating pattern $-{\rm M}-{\rm (CN)}-{\rm M}-$, where M=Au and Ag, respectively, packed on a triangular lattice to form a three-dimensional solid~\cite{sharp76}. In the former case, the chains are aligned due to the dominant metallophilic (attractive) interactions between the ${\rm Au}^+$ cations in the neighboring chains. On the other hand, in the latter case, the dominant electrostatic (repulsive) interactions between the ${\rm Ag}^+$ cations in the neighboring chains make them shift with respect to each other by 1/3 of the chain repeat length, owing to the geometrical frustration resulting from the triangular lattice geometry in the planes perpendicular to the chain direction. Thus, a unique mapping can be established by relating the relative chain shift $\Delta z$ of the cyanide structures with the phase angle $\Delta\phi$ of the XY model, through the relation $\Delta\phi=2\pi\Delta z$. While in AuCN $\Delta z=0$ maps to the ground state of the ferromagnetic $\Delta\phi=0$ XY model, in AgCN $\Delta z=\pm 1/3$ maps to the non-collinear chiral ground state of the triangular AFM XY model with $\Delta\phi=\pm 2\pi/3$.

Adding of ${\rm Ag}^+$ ions to a solution of $[{\rm Au(CN)}_2]^{-}$ in a 1:1 ratio results in a more complex bimetalic compound ${\rm Ag}_{1/2}{\rm Au}_{1/2}{\rm (CN)}$~\cite{chip12}. Such a mixed-metal system has been shown to form a line phase with strictly alternating metallic Au and Ag atoms along the chains. In spite of the existence of the intra-chain ordering the system lacks the long-range inter-chain order due to geometrical frustration arising from the effort to align chains with the preference for heterometallic (unlike) Au and Ag neighbors and the triangular geometry of the planes perpendicular to the chain direction allowing only two nearest neighbors on each elementary triangular plaquette. Cairns et al. have shown~\cite{cairns16} that the structure of such a complex bimetallic compound can be modeled by the ground states of a bilinear-biquadratic XY model~\cite{zuko03b}, which is just a special case of the present generalized XY model with $q=2$.

\begin{figure}[t!]
\subfigure{\includegraphics[scale=0.4,clip]{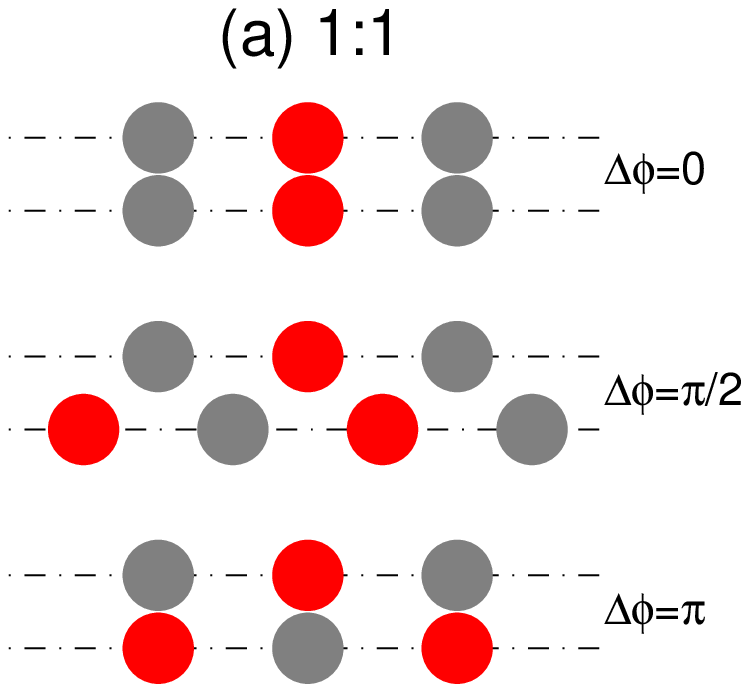}\label{fig:sch_1-1}}
\subfigure{\includegraphics[scale=0.4,clip]{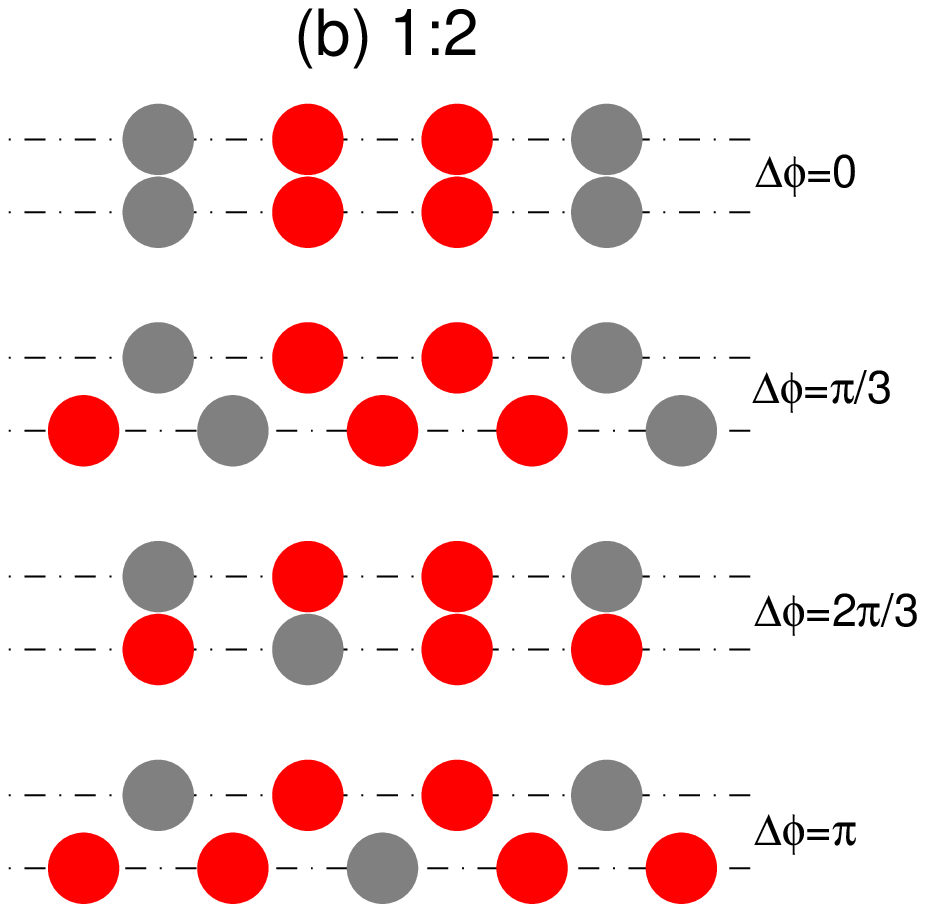}\label{fig:sch_1-2}}\\
\subfigure{\includegraphics[scale=0.4,clip]{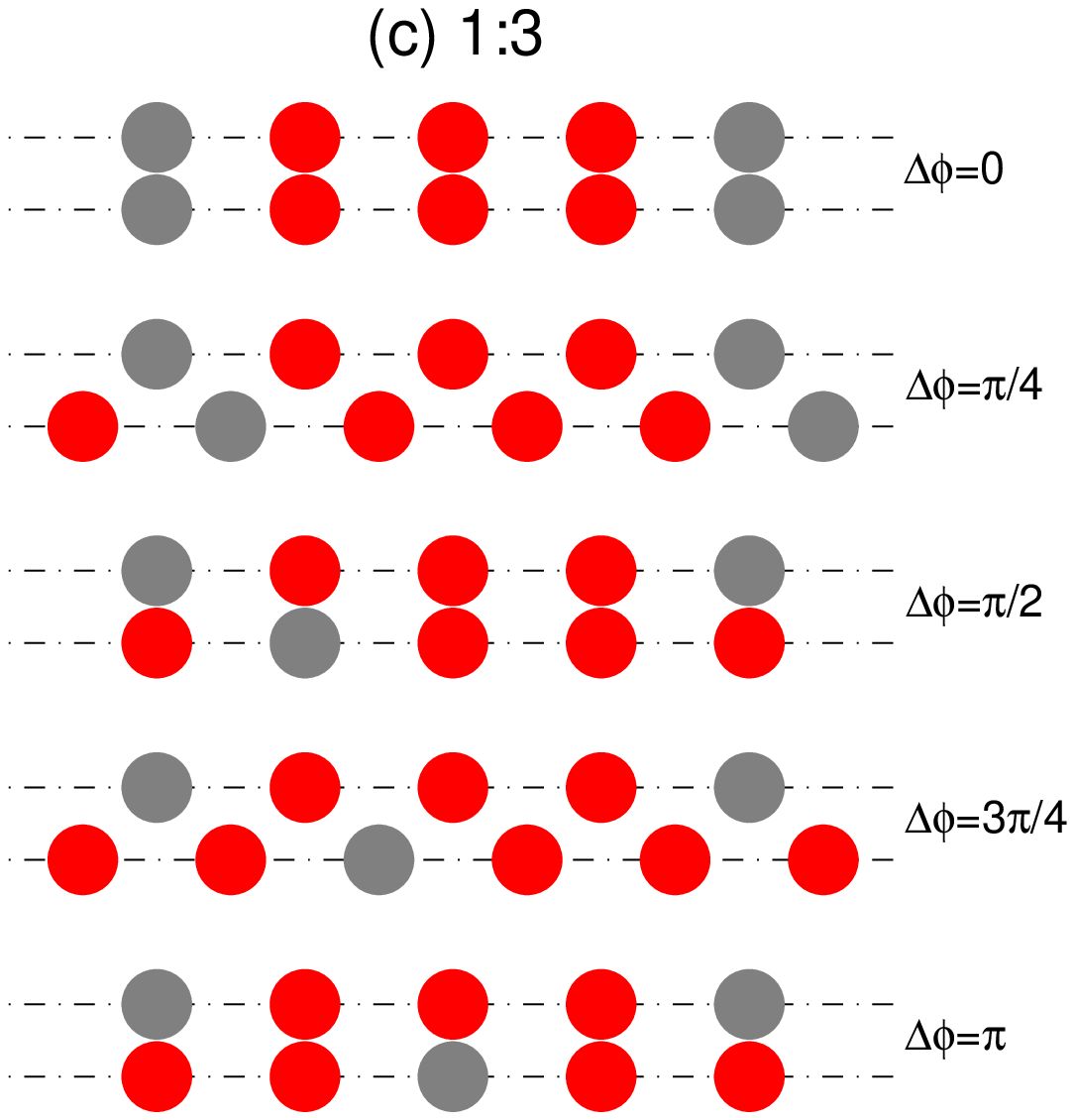}\label{fig:sch_1-3}}
\subfigure{\includegraphics[scale=0.4,clip]{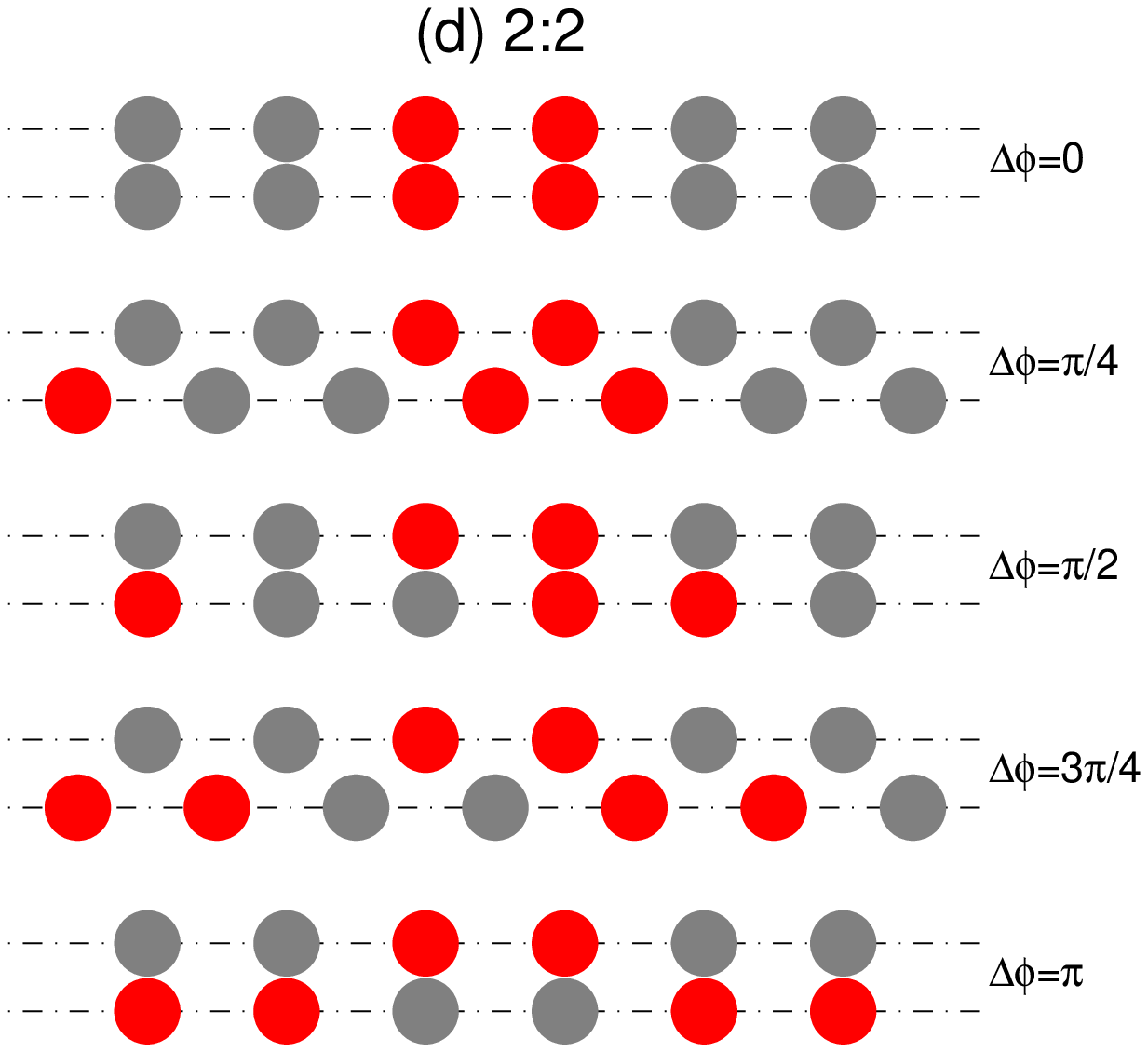}\label{fig:sch_2-2}}
\caption{(Color online) States of two neighboring metal-cyanide chains in the ${\rm M}_{x}{\rm M'}_{1-x}{\rm (CN)}$ compound, consisting of strictly-alternating linkages of metallic atoms {\rm M} (gray) and {\rm M'} (red) (${\rm CN}^-$ ions are omitted for clarity), shown for the phase shifts $\Delta\phi$ up to the half of the chain repeat length and different mixing ratios of the metallic ions.}\label{fig:map}
\end{figure} 

The mapping between the structural phases of the bimetallic compound ${\rm M}_{1/2}{\rm M'}_{1/2}{\rm (CN)}$, with a 1:1 metal-mixing ratio, such as ${\rm Ag}_{1/2}{\rm Au}_{1/2}{\rm (CN)}$, and the generalized XY model for $q=2$ is schematically shown in Fig.~\ref{fig:sch_1-1}. In particular, the cases of the phase shifts $\Delta\phi = 0$ and $\Delta\phi = \pi$ both correspond to dominant metallophilic alignments, however, in the former case the preferred alignment is homometallic while in the latter case heterometallic. On the other hand, the case of $\Delta\phi = \pi/2$ corresponds to a dominant electrostatic alignment, favoring a staggered arrangement of the metallic cations. The signs of the interaction parameters, estimated for the compound ${\rm Ag}_{1/2}{\rm Au}_{1/2}{\rm (CN)}$, suggest the preference for the heterometallic over homometallic ($J_1=-2.3(3)$ kJ/mol) and metallophilic over electrostatic ($J_2=1.3(5)$ kJ/mol) alignments. One can easily check that the interaction ratio strength corresponds to $\theta \in [\theta_{II,min},\theta_{II,max}]$. Consequently, the corresponding phase angle in Fig.~\ref{fig:gs_hs_J1_J2_q2_phi}, the order parameters in Fig.~\ref{fig:gs_J1_J2_q2_op}, as well as the snapshot in Fig.~\ref{fig:snap_q2_II} and the correlation function in Fig.~\ref{cf1_3pi_4}, indicate a complex quasi-LRO state in which magnetic disorder is coupled with (hidden) generalized nematic order.

Considering other $1:(x-1)$ metal-mixing ratios and assuming the existence of order in individual chains, such as in the gold-containing compounds, e.g., ${\rm Cu}_{2/3}{\rm Au}_{1/3}{\rm (CN)}$~\cite{chip12}, similar mappings can be established for more general bimetallic compounds ${\rm M}_{1/x}{\rm M'}_{(x-1)/x}{\rm (CN)}$, with $x>2$. In Fig.~\ref{fig:sch_1-2}, we schematically present the mapping between the structure of the ${\rm M}_{1/3}{\rm M'}_{2/3}{\rm (CN)}$ compound, with the strictly alternating chain patterns $-{\rm M}-{\rm M'}-{\rm M'}-{\rm M}-$ and the generalized XY model with $q=3$. While for the $q=2$ case the periodicity of the metallophilic/electrostatic interactions was $\Delta\phi=\pi$, for the $q=3$ case it changes to $\Delta\phi=2\pi/3$. The periodicity of the homometallic/heterometallic interactions remains the same $\Delta\phi=2\pi$. 

It is interesting to notice different roles of the geometrical frustration for the $q=2$ and $q=3$ cases with negative magnetic and generalized nematic couplings. The heterometallic pairwise interactions ($J_1<0$) are minimized by staggering neighboring chains by one half of the chain repeat length $\Delta\phi=\pi$. If only a pair of neighboring chains was considered then for $q=2$ one would see a tendency to align the metallic atoms in an alternate fashion (see the situation for $\Delta\phi=\pi$ in Fig.~\ref{fig:sch_1-1}), which would maximize the electrostatic interaction ($J_2<0$) energy and, thus, induce a competition between $J_1<0$ and $J_2<0$ interactions. On the other hand, for $q=3$ the minimum of $J_1<0$ interactions corresponds to the minimum of the $J_3<0$ interaction energy (see the situation for $\Delta\phi=\pi$ in Fig.~\ref{fig:sch_1-2}) and, thus, there is no competition between the two interactions. Nevertheless, these local arrangements cannot be propagated on the triangular lattice. For $q=2$, the lowest-energy compromise is reached by staggering neighboring chains by $\Delta\phi=\pm 2\pi/3$ and, thus, releasing the competition between $J_1<0$ and $J_2<0$, while for $q=3$, the most energetically favorable arrangement is achieved by shifting of neighboring chains by some non-universal interaction-ratio-dependent phase angle and, thus imposing competition between $J_1<0$ and $J_3<0$. This is also evident from the behavior of the phase angles and the order parameters for $\theta \in [\theta_{III,min},\theta_{III,max}]$ in Figs.~\ref{fig:gs_hs_J1_J2_q2_phi},\ref{fig:gs_J1_J2_q2_op}, for $q=2$, and Figs.~\ref{fig:gs_hs_J1_J2_q3_phi},\ref{fig:gs_J1_J2_q3_op}, for $q=3$, as well as the corresponding snapshots in Figs.~\ref{fig:snap_q2_III} and \ref{fig:snap_q3_III}.

The compound with the metal-mixing ratio 1:3 can be mapped to the $q=4$ model, with the periodicity of the metallophilic/electrostatic interactions $\Delta\phi=\pi/2$ and this can be generalized to the claim that the binary mixed-metal system with a $1:(x-1)$, $x=2,3,\ldots$, mixing ratio can be mapped to the generalized XY model with the magnetic $J_1$ and the generalized nematic $J_x$ interactions of the alignment periodicity $\Delta\phi=2\pi$ and $\Delta\phi=2\pi/x$, respectively.

Even for the same mixing ratio, one can also consider structures of different repeat patterns, such as the one shown in Fig.~\ref{fig:sch_2-2}, in which the mixing ratio 1:1 is realized by the repeat pattern of the metallic ions $-{\rm M}-{\rm M}-{\rm M'}-{\rm M'}-$ and which will be referred to as a 2:2 mixing ratio. Following the same line of arguments as in the previous paragraph, it is easy to show that such a system can be mapped to the $q=4$ model, just like the one with the 1:3 mixing ratio, and a compound with a $x:x$ mixing ratio to the generalized XY model with the magnetic $J_1$ and the generalized nematic $J_{2x}$ interactions of the alignment periodicity $\Delta\phi=2\pi$ and $\Delta\phi=\pi/x$, respectively.

\section{Conclusions}
We have studied ground state phases of a class of generalized XY models that include the standard magnetic as well as the generalized nematic higher order harmonics terms, in the model parameter space. The most intriguing are the cases when the magnetic and generalized nematic interactions induce geometrical frustration and/or mutual competition, which happens when at least one of the interactions $J_1,J_q$ is negative. Then, if they do not compete, which is the case of $J_1<0,J_q>0$ for $q$ divisible by 3 and the case of $J_1<0,J_q<0$ for $q$ non-divisible by 3, the system shows the chiral antiferromagnetic ordering, as already observed for $J_1<0,J_2<0$~\cite{park08}. Otherwise, the competition between $J_1$ and $J_q$ leads to a complex non-collinear quasi-long-range-ordered phases with still power-law decaying correlation function, as in the case of the standard XY model, however, with generally increased values of the $q$-dependent exponent $\eta$.

Furthermore, we have demonstrated that structural analogues of the generalized XY models, that are obtained by an appropriate mapping between the magnetic interactions in the XY spin models and the supramolecular interactions in the chemical systems, could be useful in structural chemistry of certain cyanide polymers. In particular, we have presented some examples of such mappings between the binary mixed-metal compounds, consisting of chains of strictly alternating patterns with different metal-mixing ratios, and the corresponding generalized XY models. One should note that in practice a perfect intra-chain metal alternation and desirable mixing ratios can be difficult to achieve but the recent experiments~\cite{chip12} suggested that, for example, the gold-containing systems of the type ${\rm M}_{p}{\rm Au}_{(1-p)}{\rm (CN)}$ might be good candidates. Structural properties of such compounds then could be predicted from the corresponding generalized XY model, using the supramolecular interactions determined by quantum mechanical calculations as the parameters $J_1$ and $J_q$.

It is evident that theoretical studies of the above models go beyond the academic interest of understanding complexity arising from the frustrated geometry and competing interactions present in the system. They can also help predict and control emergent phenomena in chemical and possibly other systems that can be viewed as structural analogues. Further extensions of the present study of the frustrated ground-state phases could include effects of thermal fluctuations, that have lead to novel phases and transitions belonging to a variety of universality classes even in the non-frustrated counterparts~\cite{pode11}, or other forms of stimuli, such as application of an external field or introduction of non-magnetic impurities, that have produces the exotic ``order by quenched disorder'' effect in the standard XY triangular-lattice antiferromagnet~\cite{mary13}. Looking for parallels in the responses to such perturbations in the magnetic systems and their chemical or other analogues can open an interesting avenue of future research.  

\begin{acknowledgments}
This work was supported by the Scientific Grant Agency of Ministry of Education of Slovak Republic (Grant No. 1/0331/15).
\end{acknowledgments}

\end{document}